\documentclass[a4paper,11pt]{article}%
\usepackage{graphicx}
\usepackage{amsbsy}
\usepackage{amsmath}
\usepackage{cite}
\usepackage{slashed}
\usepackage{epsfig}
\usepackage{amsfonts}
\usepackage{amssymb}
\usepackage{hyperref}
\hypersetup{
    colorlinks   = true,
    linkcolor    = blue,
    citecolor    = blue,
}

\begin{document}

\begin{titlepage}

\thispagestyle{empty}  \setcounter{page}{0}  \begin{flushright}

LPSC16xxx\\
December 2016\\
\end{flushright}

\vskip  2.7 true cm

\begin{center}
{\huge Towards a new paradigm for quark-lepton unification}\\[1.9cm]

\textsc{Christopher Smith}$^{\ast}$\vspace{0.5cm}\\[9pt]\smallskip
{\small \textsl{\textit{Laboratoire de Physique Subatomique et de Cosmologie,
}}}\linebreak {\small \textsl{\textit{Universit\'{e} Grenoble-Alpes,
CNRS/IN2P3, 53 avenue des Martyrs, 38026 Grenoble Cedex, France}.}}
\\[1.9cm]\textbf{Abstract}\smallskip
\end{center}

\begin{quote}
\noindent The quark and lepton mass patterns upset their na\"{\i}ve
unification. In this paper, a new approach to solve this problem is proposed.
Model-independently, we find that a successful unification can be achieved. A
mechanism is identified by which the large top quark mass renders its
third-generation leptonic partner very light. This state is thus
identified with the electron. We then provide a generic dynamical
implementation of this mechanism, using tree-level exchanges of vector leptons
to relate the quark and lepton flavor structures. In a supersymmetric context,
this same mechanism splits the squark masses, and third generation squarks end
up much lighter than the others. Finally, the implementation of this mechanism
in SU(5) GUT permits to avoid introducing any flavor structure beyond the two
minimal Yukawa couplings, ensuring the absence of unknown mixing matrices and
their potentially large impact on FCNC.

\let                                     \thefootnote          \relax
\footnotetext{$^{\ast}\;$chsmith@lpsc.in2p3.fr}
\end{quote}

\newpage%

\end{titlepage}

\tableofcontents

\section{Introduction}

Unifying all the fundamental constituents of matter has long been a major goal
of particle physics. Yet, before the advent of the Standard Model (SM), the
hadronic and leptonic particles have lived in opposite corners of our
theories. With strikingly distinct dynamics and properties, it seemed the
intimate nature of these particles were very different. This is well
illustrated by the elusive neutrinos, and the contentious conservation of
lepton number. At the same time, the much heavier protons and neutrons were
still thought to be elementary, and baryon number was, naturally, thought to be conserved.

This state of matter was of course mostly due to the strong interaction. Once
its veil is lifted, the quarks no longer seem so different from the leptons.
Their share similar weak and electromagnetic interactions, as well as the
mysterious family replication. In this sense, the SM represents the first true
milestone in their unification. As a kind of puzzling bonus, the SM also hints
at a higher level of unification. Indeed, its renormalizability, hence its
whole internal coherence, rests on the consistency between the strong and
electromagnetic charges of its fermionic constituents. In addition, baryon and
lepton numbers are not conserved in the SM, but instead the non-perturbative
electroweak interactions can for example transmute three leptons into nine
antiquarks~\cite{tHooft76}.

Soon after the SM was formulated as a spontaneously broken gauge theory, the
same receipt was used to construct Grand Unified Theories based on larger
gauge groups~\cite{GUT1,GUT2}. There, not only the interactions but also all
the matter content get embedded together in some representations of the
unified gauge group. Quarks and leptons become manifestations of the same
fundamental states, and GUT gauge interactions can transform one into the
other. These inspiring theories, however, suffer many defects yet to be
explained, most notably the stability of their scalar sector and their
prediction that the proton should decay at rates now excluded.

Whether GUT represents a true second milestone towards quark-lepton
unification is not so clear though. Indeed, embedding them in common
representations only reproduces the coherence of their strong and
electromagnetic charge we already had to impose to ensure the SM
renormalizability. This may be seen as an explanation, or as a kind of
unavoidable coincidence. Worse still, minimal GUT predicts simple relations
between quark and lepton masses, in gross disagreement with the observed
values. The only known way out of this conundrum is to somewhat relax their
unification. Disappointingly, additional Yukawa interactions have to be
introduced for the sole purpose of lifting the very prediction of unification.

The goal of this paper is to analyze the question of quark-lepton
unification. For that, in the next Section, we first take a step back from GUT
and characterize in a model independent setting the misalignment between the
quark and lepton Yukawa couplings. Our strategy is to start by assuming
\begin{equation}
\mathbf{Y}_{e}=f(\mathbf{Y}_{d},\mathbf{Y}_{u})\;,
\end{equation}
for some polynomial function $f$. Then, some requirements for a successful
unification can be deduced from the peculiarities of this function $f$, which
is found to be severely fine-tuned. In the following section, quite generic
dynamical models are constructed to alleviate this fine-tuning, and thereby
to automatically and naturally relate the quark and lepton flavor structures.
The implications of such models for supersymmetry are discussed in Section~4,
and its implementation within the minimal SU(5) model is described in Section~
5. For simplicity, neutrinos are considered massless throughout this paper.
The perspectives for neutrino mass models, as well as for other theories, are
summarized in the conclusion.

\section{Flavor symmetric perspective on quark-lepton unification\label{Sec1}}

The strategy of choice when discussing the flavor sector of any theory is to
identify the flavor symmetry and its explicit breaking terms. This permits to
systematically work out and characterize their impacts on observables. In
Section~\ref{Sec11}, we thus start by a brief summary of this technique, along
with the closely related Minimal Flavor Violation (MFV) hypothesis. This sets
the stage for Section~\ref{Sec12}, where this hypothesis is reinterpreted and
adapted to the problem at hand, which is to relate the quark and lepton Yukawa
couplings. Then, in Section~\ref{Sec13}, the peculiar fine-tuning of any
relationship between $\mathbf{Y}_{e}$, $\mathbf{Y}_{d}$, and $\mathbf{Y}_{u}$
is identified, and some generic implications for the lepton mass spectrum are
obtained. This information will guide us in the design of specific models in
Section~\ref{Sec2}.

\subsection{SM flavors and Minimal Flavor Violation\label{Sec11}}

In the SM, the three generations of matter fields can be freely and
independently redefined for each matter species without affecting the gauge
sector, which thus has the symmetry~\cite{FlavorSymm}%
\begin{equation}
G_{F}\equiv U(3)^{5}=U\left(  3\right)  _{Q}\otimes U\left(  3\right)
_{U}\otimes U\left(  3\right)  _{D}\otimes U\left(  3\right)  _{L}\otimes
U\left(  3\right)  _{E}\;,
\end{equation}
where $Q=(u_{L},d_{L})$, $U=u_{R}$, $D=d_{R}$, $L=(\nu_{L},e_{L})$, and
$E=e_{R}$. This symmetry is broken by the Yukawa couplings only, which
generate fermion masses and mixing after the electroweak symmetry breaking
(EWSB). For the following, it will prove useful to immediately generalize to a
Two Higgs Doublet Model (THDM) of type II, i.e.,%
\begin{equation}
\mathcal{L}_{Y}=\bar{U}\mathbf{Y}_{u}QH_{u}+\bar{D}\mathbf{Y}_{d}%
QH_{d}+\bar{E}\mathbf{Y}_{e}LH_{d}\;, \label{Yuk}%
\end{equation}
because then the respective normalization of the up and down quark Yukawa
couplings are tuned by the ratio of vacuum expectation values (VEV) $v_{u,d}$
of the two neutral Higgs components $H_{u,d}^{0}$, conventionally denoted as
$\tan\beta=v_{u}/v_{d}$.

As is customary, to systematically investigate the impact of these symmetry
breaking terms on observables, we first promote them to spurions. The idea is
to artificially restore the $G_{F}$ symmetry by assigning definite $G_{F}$
transformation properties to the Yukawa couplings,%
\begin{equation}
\mathbf{Y}_{u}\rightarrow g_{U}\mathbf{Y}_{u}g_{Q}^{\dagger},\;\;\mathbf{Y}%
_{d}\rightarrow g_{D}\mathbf{Y}_{d}g_{Q}^{\dagger},\;\;\mathbf{Y}%
_{e}\rightarrow g_{E}\mathbf{Y}_{e}g_{L}^{\dagger}\;,
\end{equation}
where $g_{X}\in U(3)_{X}$, so that Eq.~(\ref{Yuk}) becomes invariant under
$X\rightarrow g_{X}X$. At this stage, the SM Lagrangian becomes invariant
under $G_{F}$. Even if this is purely artificial, the amplitude for any
possible process must also be expressible as manifestly $G_{F}$-invariant, and
crucially, this may require inserting Yukawa spurions in a very specific way
in the amplitude. The $G_{F}$ symmetry thus offers a very simple tool to
predict the flavor structure of observables.

In a second stage, the spurion are frozen back to their physical values to get
quantitative predictions. The Yukawa couplings admit the Singular Value
Decompositions (SVD)%
\begin{equation}
v_{u}\mathbf{Y}_{u}=\bar{g}_{U}^{\dagger}\mathbf{m}_{u}\bar{g}_{Q}%
^{u},\;\;v_{d}\mathbf{Y}_{d}=\bar{g}_{D}^{\dagger}\mathbf{m}_{d}\bar{g}%
_{Q}^{d}\;,\;\;v_{d}\mathbf{Y}_{e}=\bar{g}_{E}^{\dagger}\mathbf{m}_{d}\bar
{g}_{L}\;,
\end{equation}
for some (fixed) $\bar{g}_{X}$ transformations. So, using the $G_{F}$
invariance, it is always possible to freeze the Yukawa couplings at the values%
\begin{equation}
v_{u}\mathbf{Y}_{u}=\mathbf{m}_{u}V,\;\;v_{d}\mathbf{Y}_{d}=\mathbf{m}%
_{d},\;\;v_{d}\mathbf{Y}_{e}=\mathbf{m}_{e}\;, \label{PhysV1}%
\end{equation}
with $\mathbf{m}_{u,d,e}=\operatorname{diag}(m_{u,d,e},m_{c,s,\mu}%
,m_{t,b,\tau})$ the diagonal mass matrices and $V=\bar{g}_{Q}^{u\dagger}%
\bar{g}_{Q}^{d}$ the CKM matrix. In this basis, the down quarks are all mass
eigenstates, but not the left-handed up quarks. Whenever convenient, the
$v_{u}\mathbf{Y}_{u}=\mathbf{m}_{u}$ and $v_{d}\mathbf{Y}_{d}=\mathbf{m}%
_{d}V^{\dagger}$ background values can also be chosen; the final results will
obviously not depend on this choice.

In the presence of New Physics (NP), assuming gauge interactions still exhibit the
$G_{F}$ symmetry, the same strategy as in the SM can be followed. In general,
there will be additional flavored couplings, which have thus to be also
promoted to spurions to restore the global $G_{F}$ symmetry. But because these
new flavor couplings are a priori generic, they could induce unacceptably
large effects in flavor observables when the New Physics scale is around the
TeV~\cite{FlavPuzzle}. On the contrary, this flavor puzzle disappears if the
hierarchies of the NP flavor couplings are similar to those observed for the
quark and lepton masses and mixings.

This is where the Minimal Flavor Violation hypothesis comes into
play~\cite{DambrosioGIS02}. It is a tool designed to systematically export the
numerical hierarchies of $\mathbf{Y}_{u,d,e}$ to the NP flavor sector, and
proceeds in two steps~\cite{MercolliS09}:

\begin{itemize}
\item Minimality: the first step is to remove the NP couplings from the
spurion list. Only $\mathbf{Y}_{u,d,e}$ are kept in order to induce the known
fermion mass. This does not forbid the NP couplings, but forces them to be
expressed as polynomial expansions in $\mathbf{Y}_{u,d,e}$, as dictated by the
$G_{F}$ symmetry.

\item Naturality: The second step requires all the free parameters to be
natural, i.e., the coefficients appearing in the spurion expansions have to be
$\mathcal{O}(1)$. This ensures that the numerical hierarchies of
$\mathbf{Y}_{u,d,e}$ are indeed passed on the NP couplings.
\end{itemize}

Provided these two conditions are met, the flavor observables are only
marginally affected by TeV NP, and the flavor puzzles are solved. We refer to
Ref.~\cite{FlavPuzzle} for more information.

\subsection{Fundamental flavor structures: Going beyond MFV\label{Sec12}}

Naively, MFV seems to treat very differently the Yukawa couplings and the NP
flavor couplings since the latter are expressed in terms of the former. For
the following, it is crucial to understand that this asymmetrical treatment of
a priori analogous Lagrangian couplings is more a matter of convenience than a
statement about their respective nature. Indeed, MFV can be interpreted as a
simple assumption about the mechanism at the origin of \textit{all} the flavor
structures~\cite{smithHDR}.

To illustrate this, imagine a low-energy theory with two elementary flavor
couplings $\mathbf{Y}$ and $\mathbf{A}$, which can be thought of as the Yukawa
and NP couplings. At the very high scale, some flavor dynamics is active and
introduces a single explicit breaking of $G_{F}$, which we call $\mathbf{X}$.
The two low-energy flavor couplings are induced by this elementary flavor
breaking, so it must be possible to relate them. For example, if $\mathbf{Y}$,
$\mathbf{A}$, and $\mathbf{X}$ all transform under the same adjoint
representation of some flavor $SU(3)\subset G_{F}$,
\begin{equation}
\left\{
\begin{array}
[c]{l}%
\mathbf{Y}=x_{1}^{Y}\mathbf{1}+x_{2}^{Y}\mathbf{X}+x_{3}^{Y}\mathbf{X}%
^{2}\;,\;\smallskip\\
\mathbf{A}=x_{1}^{A}\mathbf{1}+x_{2}^{A}\mathbf{X}+x_{3}^{A}\mathbf{X}^{2}\;.
\end{array}
\right.  \label{Redun1}%
\end{equation}
If the flavor dynamics was known, these coefficients could be computed
explicitly. Lacking this, we simply assume they are natural. Also, for these
expansions to make sense, powers of $\mathbf{X}$ must not grow unchecked. A
sufficient condition is for the trace $\langle\mathbf{X}\rangle\lesssim1$,
since then all $\mathbf{X}^{n>2}$ can be eliminated in terms of $\mathbf{1}$,
$\mathbf{X}$, and $\mathbf{X}^{2}$ without upsetting $x_{i}\sim\mathcal{O}(1)$
by using Cayley-Hamilton identities. Under this condition, from
Eq.~(\ref{Redun1}), we can get rid of the unknown high-energy spurion
$\mathbf{X}$ and derive the low-energy MFV expansions%
\begin{equation}
\left\{
\begin{array}
[c]{l}%
\mathbf{A}=y_{1}\mathbf{1}+y_{2}\mathbf{Y}+y_{3}\mathbf{Y}^{2}\;,\;\smallskip
\\
\mathbf{Y}=a_{1}\mathbf{1}+a_{2}\mathbf{A}+a_{3}\mathbf{A}^{2}\;,
\end{array}
\right.  \label{Redun2}%
\end{equation}
for some $y_{i},a_{i}$ coefficients. Naturality is preserved since
$y_{i},a_{i}\sim\mathcal{O}(1)$ when $x_{i}\sim\mathcal{O}(1)$ and
$\langle\mathbf{X}\rangle\lesssim\mathcal{O}(1)$. In practice, only the first
identity expressing $\mathbf{A}$ in terms of $\mathbf{Y}$ is useful since
$\mathbf{Y}$ is known but $\mathbf{A}$ is not. So, in this interpretation,
neither the Yukawa $\mathbf{Y}$ nor the NP coupling $\mathbf{A}$ are
fundamental, and the MFV expansions are understood as the only low-energy
observable consequences of their intrinsic redundancy.

In this paper, the basic hypothesis we wish to test is the redundancy of the
SM Yukawa couplings themselves. MFV usually assumes the minimal spurion
content to be $\mathbf{Y}_{u,d,e}$, so that all fermion masses can be induced.
Here, we want to go beyond that and express some Yukawa couplings as
expansions in others, as would happen if there are less than three fundamental
flavor couplings.

To achieve this, as a first step, we have to restrict $G_{F}$ to a smaller
group $G_{F}^{\prime}$, identify the reduced set of spurions, and fix their
transformation properties under $G_{F}^{\prime}$. There is a priori a great
latitude in these various choices and we do not plan to study them
exhaustively. Instead, with GUT settings in mind, we consider only the
continuous subgroups obtained by forcing some of the $U(3)$ transformations to
be related. In other words, from a generic transformation $(g_{U},g_{D}%
,g_{Q},g_{E},g_{L})\in G_{F}$, those of $G_{F}^{\prime}$ are obtained by
imposing the equality (modulo transpositions and/or conjugations) of some of
the $g_{i}$'s.

To further restrict the possibilities, we require

\begin{enumerate}
\item \textbf{Naturality}. With their two indices, the Yukawa couplings could
transform as $\mathbf{1}$, $\mathbf{3}$, $\mathbf{6}$, or $\mathbf{8}$ under a
given flavor $SU(3)$ or as $(\mathbf{3},\mathbf{3})$ under two different
$SU(3)$s. But, given the very hierarchical form of the Yukawa couplings,
naturality forbids any MFV expansion from starting as $\mathbf{Y}_{i}%
=a_{1}\mathbf{1}+...$, ruling out scenarios where $\mathbf{Y}_{i}%
\sim\mathbf{1}\oplus\mathbf{8}$ for some $i$. Also, if there is only one Higgs
doublets, or if $\tan\beta$ is not very large when there are two doublets,
then $G_{F}^{\prime}$ must forbid $\mathbf{Y}_{u}$ from contributing directly
to $\mathbf{Y}_{d}$ or $\mathbf{Y}_{e}$. For example, if $G_{F}^{\prime}$
allows $\mathbf{Y}_{e}=a_{1}\mathbf{Y}_{u}+...$ , then $a_{1}$ would have to
be very small.

\item \textbf{Predictivity}. When the group $G_{F}^{\prime}$ is too large
compared to the number of spurions, they can all be diagonalized and no flavor
mixing would survive. Conversely, if $G_{F}^{\prime}$ is too small compared to
the number of spurions, unknown mixing matrices render the MFV expansions
unpredictive. So $G_{F}^{\prime}$ has to give just enough freedom to rotate
all the chosen spurions to their physical background values (as is the case in
the usual MFV, see Eq.~(\ref{PhysV1})). It is then possible to bring these
spurions to their background values wherever they appear within the MFV
expansions since these are $G_{F}^{\prime}$ invariant by construction.
\end{enumerate}

In view of these points, there remain not so many viable scenarios. We need to
keep at least two spurions, the symmetry group $G_{F}^{\prime}$ has to be
large enough to account for the CKM mixing, and $\mathbf{Y}_{e,d}$ must
transform differently than $\mathbf{Y}_{u}$. The simplest choice is to
associate $\mathbf{Y}_{d}$ and $\mathbf{Y}_{e}$. For instance, if we take
\[
G_{F}^{\prime}=U(3)^{3}=U\left(  3\right)  _{Q=L}\otimes U\left(  3\right)
_{U}\otimes U\left(  3\right)  _{D=E}\;,
\]
then $\mathbf{Y}_{d}$ and $\mathbf{Y}_{e}$ transforms identically. Since only
the misalignment between $\mathbf{Y}_{u}$ and $\mathbf{Y}_{d}$ is known, and
not that between quark and lepton Yukawa couplings, the two spurions are
chosen to be%
\begin{equation}
\mathbf{Y}_{u}\rightarrow g_{U}\mathbf{Y}_{u}g_{Q}^{\dagger},\;\;\mathbf{Y}%
_{d}\rightarrow g_{D}\mathbf{Y}_{d}g_{Q}^{\dagger}\;,
\end{equation}
whose background values can be fixed as in Eq.~(\ref{PhysV1}). This pattern is
chosen also to allow for a smooth extension to GUT settings~\cite{GUTMFV}, as will
be discussed later on.

\subsection{Lepton masses from quark Yukawas\label{Sec13}}

The next step is to express $\mathbf{Y}_{e}$ as a $G_{F}^{\prime}$-symmetric
expansion in $\mathbf{Y}_{u}$ and $\mathbf{Y}_{d}$. From a mathematical point
of view, any coupling can be expressed in this way, since together with their
powers they form a complete basis for complex three-by-three
matrices~\cite{ColangeloNS08,Nikolidakis08}. What matters is the size of the
expansion coefficients. Generic matrices expanded in such a basis require huge
coefficients, while we are after $\mathcal{O}(1)$ ones for naturality reasons.

To illustrate this, consider the most general expansion, given the
$G_{F}^{\prime}$ properties,%
\begin{align}
\mathbf{Y}_{e}  &  =c_{0}\mathbf{Y}_{d}\cdot(\mathbf{1}+c_{1}\mathbf{Y}%
_{u}^{\dagger}\mathbf{Y}_{u}+c_{2}\mathbf{Y}_{d}^{\dagger}\mathbf{Y}_{d}%
+c_{3}(\mathbf{Y}_{u}^{\dagger}\mathbf{Y}_{u})^{2}+c_{4}(\mathbf{Y}%
_{d}^{\dagger}\mathbf{Y}_{d})^{2}+c_{5}\{\mathbf{Y}_{u}^{\dagger}%
\mathbf{Y}_{u},\mathbf{Y}_{d}^{\dagger}\mathbf{Y}_{d}\}\nonumber\\
&  \;\;\;\;\;\;\;\;\;\;\;\;\;\;\;\;+ic_{6}[\mathbf{Y}_{u}^{\dagger}%
\mathbf{Y}_{u},\mathbf{Y}_{d}^{\dagger}\mathbf{Y}_{d}]+ic_{7}[(\mathbf{Y}%
_{u}^{\dagger}\mathbf{Y}_{u})^{2},\mathbf{Y}_{d}^{\dagger}\mathbf{Y}%
_{d}]+ic_{8}[\mathbf{Y}_{u}^{\dagger}\mathbf{Y}_{u},(\mathbf{Y}_{d}^{\dagger
}\mathbf{Y}_{d})^{2}])\;. \label{BasicExp}%
\end{align}
If we require that this equation holds exactly once $\mathbf{Y}_{u,d,e}$ are
replaced by their background values Eq.~(\ref{PhysV1}), then only terms
involving $\mathbf{Y}_{d}$ can contribute since $\mathbf{Y}_{u}$ is not
diagonal. The equation can nevertheless be solved but huge coefficients are
required%
\begin{equation}
c_{0}=0.2\;,\;\;c_{2}=7\times10^{7}\times r_{\beta}^{2}\;,\;\;c_{4}%
=-3\times10^{11}\times r_{\beta}^{4}\;,\;\;c_{i\neq2,4}=0\;, \label{DiagSVD}%
\end{equation}
where $r_{\beta}=50/\tan\beta$ encodes a simplified $\tan\beta$ scaling, valid
for $\tan\beta\gtrsim5$. This is way beyond natural, but sets the stage
against which we can compare more realistic settings. Also, it serves to
illustrate how sensitive the coefficients are when trying to fit even slight misalignments.

Of course, it makes no sense to require $\mathbf{Y}_{e}$ to be diagonal in the
basis in which $v_{u}\mathbf{Y}_{u}=\mathbf{m}_{u}V$ and $v_{d}\mathbf{Y}%
_{d}=\mathbf{m}_{d}$. Once the consequence of $G_{F}^{\prime}$ are worked out,
the leptons are free to be rotated independently of the quarks. So, all that
is required is for the three singular values of $v_{d}\mathbf{Y}_{e}$ to match
the observed lepton masses. This means that there are only three constraints
to solve for the nine a priori complex coefficients, leaving a large
under-determination. To cure for this, we start by keeping only the three
simplest terms in the expansion and set%
\begin{equation}
c_{i\geq3}\equiv0\;.
\end{equation}
Restricting coefficients to real values, and using the fermion masses quoted
in Ref.~\cite{Bora} for several scenarios, we find%
\begin{equation}%
\begin{tabular}
[c]{rlll}%
Masses at$\text{\thinspace}M_{Z}:\;\;$ & $\;c_{0}=8.6\;,\;$ & $\;\;c_{1}%
=-1.8\;,\;$ & $\;\;c_{2}=1.2\times r_{\beta}^{2}\;,\;$\\
$\text{SM at\thinspace}M_{GUT}:\;\;$ & $\;c_{0}=22\;,$ & $\;\;c_{1}%
=\;\;6\;,\;$ & $\;\;c_{2}=-5\times10^{4}\;,$\\
$\text{MSSM at\thinspace}M_{GUT}:\;\;$ & $\;c_{0}=20\;,$ & $\;\;c_{1}%
=-7.9\;,\;$ & $\;\;c_{2}=\;\;5.3\times r_{\beta}^{2}\;,$\\
THDM$\text{ at\thinspace}M_{GUT}:\;\;$ & $\;c_{0}=20\;,$ & $\;\;c_{1}%
=-8.6\;,\;$ & $\;\;c_{2}=\;\;5.0\times r_{\beta}^{2}\;.$%
\end{tabular}
\ \ \ \ \ \ \ \ \label{CoefPoly}%
\end{equation}
The sign of $c_{0}$ is not fixed since it is irrelevant for the SVD values.
Allowing for all the terms of Eq.~(\ref{BasicExp}) permits to reduce $c_{1,2}$
a bit but does not change their order of magnitude.

It is truly remarkable that it is possible for at least some of the scenarios
to obtain natural values for the expansion coefficients. The most natural
values arise at the EW scale, when $\tan\beta$ is sufficiently large to make
$\mathbf{Y}_{d}^{\dagger}\mathbf{Y}_{d}$ entries of comparable size to those
of $\mathbf{Y}_{u}^{\dagger}\mathbf{Y}_{u}$. Beyond that scale, the RG
evolution under the MSSM or THDM at moderate or high $\tan\beta$ is strongly
favored, while that of the SM departs from naturality essentially because $\tan
\beta=1$, and also because the specific hierarchies of the Yukawa couplings
becomes less compatible.

\subsubsection{On the anatomy of a fine-tuning}

\begin{figure}[t]
\centering     \includegraphics[width=0.9\textwidth]{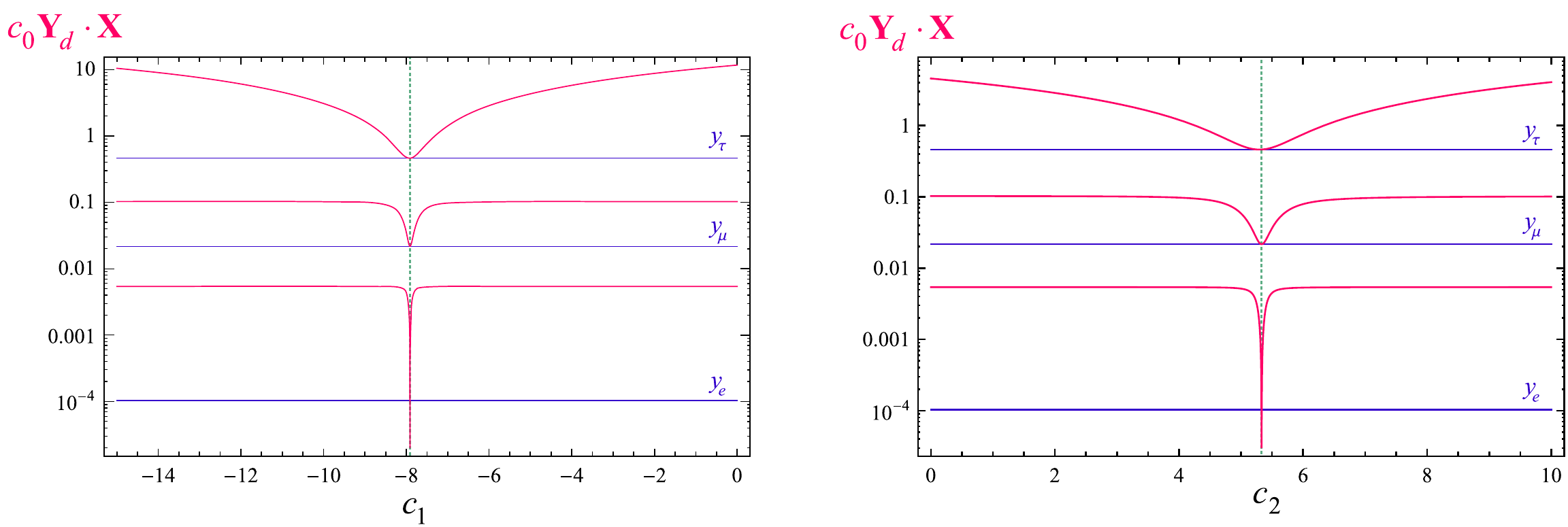}
\caption{Evolution of the singular values of $\mathbf{Y}_{e}=c_{0}\mathbf{Y}_{d}\cdot\mathbf{X}$, with $\mathbf{X}\equiv\mathbf{1}+c_{1}\mathbf{Y}_{u}^{\dagger}\mathbf{Y}_{u} +c_{2}\mathbf{Y}_{d}^{\dagger}\mathbf{Y}_{d}$, as a function of the coefficient $c_1$ or $c_2$, holding the other parameters fixed. This evolution is very smooth, except for the dip occurring when $c_1$ or $c_2$ is such that $\det\mathbf{X} = 0$. The green dashed lines indicate the point at which the singular values of $\mathbf{Y}_{e}$ coincide with the observed lepton Yukawa couplings, which clearly sits deep inside the dip (these values correspond to the MSSM at the GUT scale and $\tan \beta = 50$, see Eq.~(\ref{CoefPoly})).}
\label{Fig1}
\end{figure}

The size of the coefficients is not the only measure of naturalness. Despite
their reasonable appearance, these expansions are severely fined-tuned. The
behavior of the singular values when one of the expansion parameters is
allowed to vary is shown in Fig.~\ref{Fig1}. Clearly, the polynomial expansion with
natural coefficients has a marginal effect and the singular values stay very
close to those of $c_{0}\times\mathbf{Y}_{d}$ except for a peculiar point
where they all suddenly dip. If we denote the polynomial%
\begin{equation}
\mathbf{X}\equiv\mathbf{1}+c_{1}\mathbf{Y}_{u}^{\dagger}\mathbf{Y}_{u}%
+c_{2}\mathbf{Y}_{d}^{\dagger}\mathbf{Y}_{d}\;, \label{FTcancel0}%
\end{equation}
so that $\mathbf{Y}_{e}=c_{0}\mathbf{Y}_{d}\cdot\mathbf{X}$, what happens at
that point is a near cancellation%
\begin{equation}
1\approx\mathbf{X}^{11}\approx\mathbf{X}^{22}\gg\mathbf{X}^{33}\approx0\;.
\label{FTcancel}%
\end{equation}
For example, in the MSSM at $\tan\beta=50$,%
\begin{equation}
|\mathbf{X}|=\left(
\begin{array}
[c]{ccc}%
1 & 0.0005 & 0.01\\
0.0005 & 1 & 0.06\\
0.01 & 0.06 & 0.004
\end{array}
\right)  \;. \label{InvertX}%
\end{equation}
The eigenvalues of this polynomial show an even more striking hierarchy, with
$v_{1}=1.0017$, $v_{2}=1.0000$, but $v_{3}=0.00026$. It is this peculiar
feature which permits to significantly twist the singular values of
$\mathbf{Y}_{d}$ to reproduce those of $\mathbf{Y}_{e}$.

To better understand why, in the basis Eq.~(\ref{PhysV1}), the $33$ entry seem
to play a particular role, let us take the determinant of Eq.~(\ref{BasicExp}%
). The unknown SVD matrices are unitary and disappear, leaving%
\begin{equation}
\det(\mathbf{Y}_{e})=c_{0}^{3}\times\det(\mathbf{Y}_{d})\times\det
(\mathbf{1}+c_{1}\mathbf{Y}_{u}^{\dagger}\mathbf{Y}_{u}+c_{2}\mathbf{Y}%
_{d}^{\dagger}\mathbf{Y}_{d})\;.
\end{equation}
The dip shown in Fig.~\ref{Fig1} corresponds to the point where $\det\mathbf{X}$
vanishes. Using Cayley-Hamilton identities and thanks to the large hierarchy
of the Yukawa couplings, $(\mathbf{Y}_{i}^{\dagger}\mathbf{Y}_{i})^{2}%
\approx\langle\mathbf{Y}_{i}^{\dagger}\mathbf{Y}_{i}\rangle\mathbf{Y}%
_{i}^{\dagger}\mathbf{Y}_{i}$, $i = u,d$,%
\begin{equation}
\det(\mathbf{1}+c_{1}\mathbf{Y}_{u}^{\dagger}\mathbf{Y}_{u}+c_{2}%
\mathbf{Y}_{d}^{\dagger}\mathbf{Y}_{d})\approx1+c_{1}\langle\mathbf{Y}%
_{u}^{\dagger}\mathbf{Y}_{u}\rangle+c_{2}\langle\mathbf{Y}_{d}^{\dagger
}\mathbf{Y}_{d}\rangle\approx0\;. \label{FTcancelDET}%
\end{equation}
In the basis Eq.~(\ref{PhysV1}), this immediately implies Eq.~(\ref{FTcancel})
since the top and bottom Yukawa couplings dominate, $\langle\mathbf{Y}%
_{u,d}^{\dagger}\mathbf{Y}_{u,d}\rangle\approx y_{t,b}^{2}$ . The fact that
natural coefficients are possible at all can thus be traced to the large
$y_{t,b}^{2}$ couplings. In this sense, it looks truly remarkable that a
solution where both $c_{1}$ and $c_{2}$ end up not larger than $\langle
\mathbf{Y}_{u}^{\dagger}\mathbf{Y}_{u}\rangle^{-1}$ and $\langle\mathbf{Y}%
_{d}^{\dagger}\mathbf{Y}_{d}\rangle^{-1}$ exists. Still, at this stage, we
cannot make the economy of a mechanism able to automatically ensure such a
near cancellation of $\det(\mathbf{X})$.

As a side remark, it should be noted that solving Eq.~(\ref{BasicExp}) for $c_{0,1,2}$ given the singular values of $\mathbf{Y}_{e}$ is tricky. Indeed, singular value decompositions are highly non-linear, and the equations for $c_{0,1,2}$ cannot be solved exactly. Worse, once reverting to numerical methods, algorithms are very unstable because the solutions we are after lie in the very narrow valley where the required cancellation takes place.

\subsubsection{The twisted persona of the leptons}

Before turning to scenarios, there is another peculiar feature of the
expansion worth discussing. The SVD of $\mathbf{Y}_{e}$ is $g_{E}%
\mathbf{Y}_{e}g_{L}^{\dagger}$, so let us look at the mixing matrices $g_{E}$
and $g_{L}^{\dagger}$ as one approaches the dip of Fig.~\ref{Fig1}. We thus take the MSSM
at $\tan\beta=50$ and vary $c_{2}$ holding the other coefficients $c_{0,1}$
fixed. Away from the dip, the two unitary matrices deviate only slightly from
identity%
\begin{equation}
c_{2}=2.5:|g_{E}|=\left(
\begin{array}
[c]{ccc}%
1.000 & 0.00016 & 0.000054\\
0.00016 & 1.000 & 0.0048\\
0.000055 & 0.0048 & 1.000
\end{array}
\right)  \;,\;\;|g_{L}|=\left(
\begin{array}
[c]{ccc}%
1.00 & 0.0050 & 0.038\\
0.0019 & 0.98 & 0.18\\
0.038 & 0.18 & 0.98
\end{array}
\right)  \;.
\end{equation}
Moving closer, the situation changes dramatically for $g_{L}$%
\begin{equation}
c_{2}=4.8:|g_{E}|=\left(
\begin{array}
[c]{ccc}%
1.000 & 0.0084 & 0.0012\\
0.0082 & 0.99 & 0.11\\
0.0020 & 0.11 & 0.99
\end{array}
\right)  \;,\;\;|g_{L}|=\left(
\begin{array}
[c]{ccc}%
0.98 & 0.13 & 0.14\\
0.011 & 0.71 & 0.70\\
0.20 & 0.69 & 0.70
\end{array}
\right)  \;.
\end{equation}
The reason for this large mixing in the left-handed lepton sector is the
difference between
\begin{equation}
\mathbf{Y}_{e}^{\dagger}\mathbf{Y}_{e}=c_{0}^{2}\times\mathbf{X}%
\cdot\mathbf{Y}_{d}^{\dagger}\mathbf{Y}_{d}\cdot\mathbf{X}\;\text{
and}\;\;\;\mathbf{Y}_{e}\mathbf{Y}_{e}^{\dagger}=c_{0}^{2}\times\mathbf{Y}%
_{d}\cdot\mathbf{X}^{2}\cdot\mathbf{Y}_{d}^{\dagger}\;,
\end{equation}
diagonalized by the unitary matrix $g_{L}$ and $g_{E}$, respectively. Because
of Eq.~(\ref{FTcancel}), the entry $(\mathbf{Y}_{e}^{\dagger}\mathbf{Y}%
_{e})^{33}$ decreases approaching the dip, but this does not occur for
$\mathbf{Y}_{e}\mathbf{Y}_{e}^{\dagger}$ whose diagonal entries always stay
very hierarchical. The point $c_{2}=4.8$ corresponds to $(\mathbf{Y}%
_{e}^{\dagger}\mathbf{Y}_{e})^{33}\approx(\mathbf{Y}_{e}^{\dagger}%
\mathbf{Y}_{e})^{22}$, hence the large mixing present in $g_{L}$.

Moving even closer to the dip, $(\mathbf{Y}_{e}^{\dagger}\mathbf{Y}_{e})^{33}$
becomes smaller than $(\mathbf{Y}_{e}^{\dagger}\mathbf{Y}_{e})^{22}$ and the
left-handed leptons get even more twisted:
\begin{equation}
c_{2}=5.2:|g_{E}|=\left(
\begin{array}
[c]{ccc}%
0.97 & 0.086 & 0.0023\\
0.083 & 0.98 & 0.20\\
0.020 & 0.20 & 0.98
\end{array}
\right)  \;,\;\;|g_{L}|=\left(
\begin{array}
[c]{ccc}%
0.80 & 0.56 & 0.20\\
0.033 & 0.28 & 0.96\\
0.59 & 0.77 & 0.21
\end{array}
\right)  \;.
\end{equation}
At the $c_{2}$ value for which $g_{E}\mathbf{Y}_{e}g_{L}^{\dagger}%
=\mathbf{m}_{e}/v_{d}$, the mixings settle at
\begin{equation}
c_{2}=5.3:|g_{E}|=\left(
\begin{array}
[c]{ccc}%
0.97 & 0.24 & 0.0023\\
0.24 & 0.95 & 0.22\\
0.055 & 0.21 & 0.98
\end{array}
\right)  \;,\;\;|g_{L}|=\left(
\begin{array}
[c]{ccc}%
0.032 & 0.98 & 0.20\\
0.062 & 0.20 & 0.98\\
1.00 & 0.019 & 0.068
\end{array}
\right)  \;. \label{TwistPoly}%
\end{equation}
At this value, large mixing angles disappear and all mixings are CKM-like.
Still, the left-handed leptons are irremediably twisted since%
\begin{equation}
\left(
\begin{array}
[c]{c}%
e_{L}\\
\mu_{L}\\
\tau_{L}%
\end{array}
\right)  ^{phys}\approx\left(
\begin{array}
[c]{ccc}%
0 & 0 & 1\\
1 & 0 & 0\\
0 & 1 & 0
\end{array}
\right)  \cdot\left(
\begin{array}
[c]{c}%
e_{L}\\
\mu_{L}\\
\tau_{L}%
\end{array}
\right)  ^{gauge}\;.
\end{equation}
Note that this reordering of the leptonic states does not depend on the basis
chosen for the quark Yukawa couplings in Eq.~(\ref{PhysV1}), contrary to the
mixing angles in $g_{E}$ and $g_{L}$. In practice, as long as neutrinos are
massless and in the absence of lepton-number violating couplings, neither
these mixings nor the twist are observable. On the other hand, when studying
the neutrino sector, especially mass hierarchies, such a twist could have great
implications since the lightest left handed lepton would be essentially the
third-generation gauge state.

As a final remark, it should be noted that the results of this section do not
change if one identifies the flavor group as $U\left(  3\right)  _{Q=E}\otimes
U\left(  3\right)  _{U}\otimes U\left(  3\right)  _{D=L}$ instead of $U\left(
3\right)  _{Q=L}\otimes U\left(  3\right)  _{U}\otimes U\left(  3\right)
_{D=E}$, except for the interchange of $g_{L}$ and $g_{E}$. Indeed, the SVD
constraints imposing $\mathbf{Y}_{e}=c_{0}\mathbf{Y}_{d}\cdot\mathbf{X}$ or
$\mathbf{Y}_{e}^{T}=c_{0}\mathbf{Y}_{d}\cdot\mathbf{X}$ are obviously
identical, but for $g_{L}\leftrightarrow g_{E}$. The right-handed leptons
would then be twisted, with no visible consequence on the neutrinos. Further,
we will see in the next section that it is also possible to have double
expansions like $\mathbf{Y}_{e}=c_{0}\mathbf{X\cdot Y}_{d}\cdot\mathbf{X}%
^{\prime}$ with both $\det\mathbf{X}$ and $\det\mathbf{X}^{\prime}$ close to
zero, in which case right and left leptons end up simultaneously twisted. This
should be kept in mind, especially as the $SU(5)$ unification pattern
corresponds~\cite{GUTMFV} to $U\left(  3\right)  _{Q=E=U}\otimes U\left(
3\right)  _{D=L}$, with $\mathbf{Y}_{u}\sim(\mathbf{\bar{6}},\mathbf{1})$
required to be symmetric, and $\mathbf{Y}_{d}\sim\mathbf{Y}_{e}^{T}%
\sim(\mathbf{\bar{3}},\mathbf{3})$.

\section{Scenario 1: Light electrons from heavy tops\label{Sec2}}

It is now time to devise a mechanism able to naturally tune the MFV expansion
of $\mathbf{Y}_{e}$. In the next subsection, this problem is tackled from a
mathematical point of view, and in the following, a corresponding physically
plausible though quite generic scenario is presented.

\subsection{The mathematics of infinite MFV expansions}

Let us restate the problem at hand. We have seen that the expansion
$\mathbf{Y}_{e}=c_{0}\mathbf{Y}_{d}\cdot\mathbf{X}$ requires $\mathbf{X}%
^{33}\approx0$. This means, dropping $\mathbf{Y}_{d}^{\dagger}\mathbf{Y}_{d}$
for simplicity, that with%
\begin{equation}
\mathbf{X}=\mathbf{1}+c\mathbf{Y}_{u}^{\dagger}\mathbf{Y}_{u}\;, \label{Sol1}%
\end{equation}
the coefficient must be tuned to%
\begin{equation}
\det\mathbf{X}\approx0\Rightarrow c\approx-\frac{1}{\langle\mathbf{Y}%
_{u}^{\dagger}\mathbf{Y}_{u}\rangle}\;. \label{Sol2}%
\end{equation}
Though the numerical value of $c$ is natural thanks to the large top quark
Yukawa coupling, the fine-tuning between $c$ and $\langle\mathbf{Y}%
_{u}^{\dagger}\mathbf{Y}_{u}\rangle$ is unacceptable. Clearly, adding more
terms to the $\mathbf{X}$ expansion cannot improve the situation. For example,
if we add a term $c^{\prime}(\mathbf{Y}_{u}^{\dagger}\mathbf{Y}_{u})^{2}$ to
$\mathbf{X}$, then both $c$ and $c^{\prime}$ have to be fined-tuned so that
$\det\mathbf{X}\approx0$. No finite polynomial in $\mathbf{Y}_{u}^{\dagger
}\mathbf{Y}_{u}$ and/or $\mathbf{Y}_{d}^{\dagger}\mathbf{Y}_{d}$ would ever
permit to relax the fine-tuning.

The key to solve this problem is to consider infinite polynomials. Consider
for instance the geometric series%
\begin{equation}
\mathbf{X}=\mathbf{1}+\eta\mathbf{Y}_{u}^{\dagger}\mathbf{Y}_{u}+\eta
^{2}(\mathbf{Y}_{u}^{\dagger}\mathbf{Y}_{u})^{2}+\eta^{3}(\mathbf{Y}%
_{u}^{\dagger}\mathbf{Y}_{u})^{3}+...
\end{equation}
Barring convergence issues to be discussed below, the sum is%
\begin{equation}
\mathbf{X}=\frac{1}{\mathbf{1}-\eta\mathbf{Y}_{u}^{\dagger}\mathbf{Y}_{u}}\;.
\label{resummX}%
\end{equation}
This matrix has the desired property. In the diagonal basis, $\mathbf{Y}%
_{u}^{\dagger}\mathbf{Y}_{u}=\operatorname{diag}(y_{u}^{2},y_{c}^{2},y_{t}%
^{2})$ and%
\begin{equation}
\mathbf{X}^{11,22}=\frac{1}{1-\eta y_{u,c}^{2}}\approx1\;\;\text{but
\ }\mathbf{X}^{33}=\frac{1}{1-\eta y_{t}^{2}}\approx0\;,
\end{equation}
whenever $\eta$ is large enough that $\eta y_{t}^{2}\gg1$ but still small
enough that $\eta y_{u,c}^{2}\ll1$. Specifically, the large top quark mass
translate into $(\mathbf{Y}_{u}^{\dagger}\mathbf{Y}_{u})^{2}\approx
\langle\mathbf{Y}_{u}^{\dagger}\mathbf{Y}_{u}\rangle\mathbf{Y}_{u}^{\dagger
}\mathbf{Y}_{u}$, so that%
\begin{equation}
\mathbf{X=}\sum_{n=0}^{\infty}\eta^{n}(\mathbf{Y}_{u}^{\dagger}\mathbf{Y}%
_{u})^{n}\approx\mathbf{1}+\eta\mathbf{Y}_{u}^{\dagger}\mathbf{Y}_{u}%
\sum_{n=0}^{\infty}\eta^{n}\langle\mathbf{Y}_{u}^{\dagger}\mathbf{Y}%
_{u}\rangle^{n}=\mathbf{1}+\frac{\eta}{1-\eta\langle\mathbf{Y}_{u}^{\dagger
}\mathbf{Y}_{u}\rangle}\mathbf{Y}_{u}^{\dagger}\mathbf{Y}_{u}\;,
\end{equation}
which tends to%
\begin{equation}
\mathbf{X}\overset{\eta\gg1}{\approx}\mathbf{1}-\frac{1}{\langle\mathbf{Y}%
_{u}^{\dagger}\mathbf{Y}_{u}\rangle}\mathbf{Y}_{u}^{\dagger}\mathbf{Y}_{u}\;.
\label{Resummed3}%
\end{equation}
This is precisely the result we were after, Eqs.~(\ref{Sol1}) and~(\ref{Sol2}%
). Crucially, the value of $\eta$ does not need to have any precise
relationship with $\langle\mathbf{Y}_{u}^{\dagger}\mathbf{Y}_{u}\rangle$, it
just needs to be large enough so that $\eta\langle\mathbf{Y}_{u}^{\dagger
}\mathbf{Y}_{u}\rangle\gg1$.

Evidently, the suppression of $\mathbf{X}^{33}$ requires summing the geometric
series well outside its radius of convergence. Even if one could argue that
such series make sense through analytic continuation, as is customary for
perturbative series in Quantum Field Theory, the situation is not very
comfortable. One simple way out of possible convergence issues is to consider
for example Eq.~(\ref{resummX}) as the true expression. In this way, even if
the expanded form of the MFV polynomial does not converge from a strict
mathematical sense, it should not have been trusted in the first place. We
will see in the next section a practical realization of such a scenario. One
should note also a peculiar feature of the geometric series involving
matrices. Even if the infinite sum of powers does not converge, any inverse
matrix can be expanded in a finite polynomial. Denoting $\mathbf{A}\equiv
\eta\mathbf{Y}_{u}^{\dagger}\mathbf{Y}_{u}$ and using Cayley-Hamilton
identities,%
\begin{equation}
\frac{1}{\mathbf{1}+\mathbf{A}}=\frac{1}{\det(\mathbf{1}+\mathbf{A})}\left[
\mathbf{1(}1+\langle\mathbf{A}\rangle+\frac{1}{2}(\langle\mathbf{A}\rangle
^{2}-\langle\mathbf{A}^{2}\rangle))-\mathbf{A}\left(  1+\langle\mathbf{A}%
\rangle\right)  +\mathbf{A}^{2}\right]  \;,
\end{equation}
whenever
\begin{equation}
\det(\mathbf{1}+\mathbf{A})=1+\langle\mathbf{A}\rangle+\frac{1}{2}%
\langle\mathbf{A}\rangle^{2}-\frac{1}{2}(1+\langle\mathbf{A}\rangle
)\langle\mathbf{A}^{2}\rangle+\frac{1}{3}\langle\mathbf{A}^{3}\rangle
+\frac{1}{6}\langle\mathbf{A}\rangle^{3}\neq0\;.
\end{equation}
The result Eq.~(\ref{Resummed3}) is immediately obtained in the third
generation dominance $\mathbf{A}^{2}\rightarrow\langle\mathbf{A}%
\rangle\mathbf{A}$, even though no resummation is implied.

All the discussions of this section can be extended to include both
$\mathbf{A}\equiv\eta\mathbf{Y}_{u}^{\dagger}\mathbf{Y}_{u}$ and
$\mathbf{B}\equiv\eta^{\prime}\mathbf{Y}_{d}^{\dagger}\mathbf{Y}_{d}$. The
analytical expressions are more cumbersome since in general $[\mathbf{A}%
,\mathbf{B}]\neq0$. For example,%
\begin{equation}
\mathbf{X}=(\mathbf{1}+\mathbf{A}+\mathbf{A}^{2}+...)(\mathbf{1}%
+\mathbf{B}+\mathbf{B}^{2}+...)=\frac{1}{\mathbf{1}+\mathbf{A}}\frac{1}%
{\mathbf{1}+\mathbf{B}}\approx\left(  \mathbf{1}-\frac{\mathbf{A}}%
{1+\langle\mathbf{A}\rangle}\right)  \left(  \mathbf{1}-\frac{\mathbf{B}%
}{1+\langle\mathbf{B}\rangle}\right)  \;,
\end{equation}
where the last equality holds in the third-generation dominance approximation,
or%
\begin{equation}
\mathbf{X}=\mathbf{1}+\mathbf{A}+\mathbf{B}+(\mathbf{A}+\mathbf{B}%
)^{2}+...=\frac{1}{\mathbf{1}+\mathbf{A}+\mathbf{B}}\approx\mathbf{1}%
-\frac{\mathbf{A}+\mathbf{B}}{1+\langle\mathbf{A}\rangle+\langle
\mathbf{B}\rangle}\;.
\end{equation}
Both these series manifestly\footnote{Care is needed though when
simultaneously working in the third-generation dominance approximation and
performing the $\eta\rightarrow\infty$ limit, as the latter is not fully
compatible with the former.} reproduces the previous result $\mathbf{X}%
^{11},\mathbf{X}^{22}\gg\mathbf{X}^{33}\approx0$ thanks to the large hierarchy
in the $\mathbf{Y}_{u,d}$ couplings, and require analytical continuation to
be defined outside of their radius of convergence.

\subsection{Vector-like leptons and geometric Yukawas\label{SecVL}}

To induce geometric-like MFV expansions, our strategy is to generate effective
contributions to the Yukawa coupling $\mathbf{Y}_{e}$ through the tree-level
exchange of new states. As such, it is a bit similar to the Froggatt-Nielsen
mechanism~\cite{Froggatt:1978nt}, although the new fields will not introduce
any new breaking of the flavor symmetry. Such breaking would not be adequate
here since the goal is to generate the MFV series, not to explain the internal
hierarchy of the Yukawa couplings themselves.

Specifically, consider adding to the SM a flavor-triplet of vector leptons
$X_{L,R}$, having the same gauge quantum numbers as the lepton doublet, and a
singlet scalar boson $H_{s}$. The new terms in the Lagrangian are, omitting
flavor indices for simplicity%
\begin{align}
\mathcal{L}_{X_{L,R}}  &  =\bar{X}_{L}(i \slashed{D})X_{L}+\bar{X}_{R}(i \slashed{D}))X_{R}+\bar{X}_{L}\mathbf{M}_{X}X_{R}+\bar{X}_{R}\mathbf{M}_{X}X_{L}\frac{{}}{{}}\nonumber\\
&  +\left(  \bar{X}_{R}\mathbf{N}_{X}X_{L}H_{s}+\bar{X}_{R}\mathbf{\bar{N}%
}_{X}LH_{s}+\bar{E}\mathbf{Y}_{X}X_{L}H_{d}+h.c.\frac{{}}{{}}\right)  \;,
\label{LagrX}%
\end{align}
where $\mathbf{M}_{X}$, $\mathbf{N}_{X}$, $\mathbf{\bar{N}}_{X}$ and
$\mathbf{Y}_{X}$ are all three-by-three matrices in flavor space. This model
contains many new flavor couplings and flavored particles, so our starting
point is to impose MFV. For that, we take the flavor symmetry%
\begin{equation}
G_{F}^{\prime}=U\left(  3\right)  _{Q=L=X_{L,R}}\otimes U\left(  3\right)
_{U}\otimes U\left(  3\right)  _{D=E}\;,
\end{equation}
with thus $X_{L,R}$ transforming like $L$ and $Q$, and allow only for
$\mathbf{Y}_{u}$ and $\mathbf{Y}_{d}$ as spurions. The various flavor
couplings can then all be expressed in terms of $\mathbf{Y}_{u}$ and
$\mathbf{Y}_{d}$. We assume the simple expansions%
\begin{equation}
\left\{
\begin{array}
[c]{c}%
\mathbf{M}_{X}=M_{X}\mathbf{1\;,}\\
\mathbf{Y}_{e}=\mathbf{Y}_{X}=\gamma\mathbf{Y}_{d}\;,\\
\mathbf{N}_{X}=\mathbf{\bar{N}}_{X}=\alpha\mathbf{Y}_{u}^{\dagger}%
\mathbf{Y}_{u}+\beta\mathbf{Y}_{d}^{\dagger}\mathbf{Y}_{d}\;,
\end{array}
\right.  \label{CondX}%
\end{equation}
where $\mathbf{Y}_{e}$ corresponds to the SM Yukawa interaction $E\mathbf{Y}%
_{e}LH_{d}$. A constant term in $\mathbf{N}_{X}=\mathbf{\bar{N}}_{X}$ is
omitted even if it is consistent with $G_{F}^{\prime}$ for reasons that will
be clear below, so we assume that these couplings disappear in the absence of
$\mathbf{Y}_{u,d}$.

\begin{figure}[t]
\centering     \includegraphics[width=0.8\textwidth]{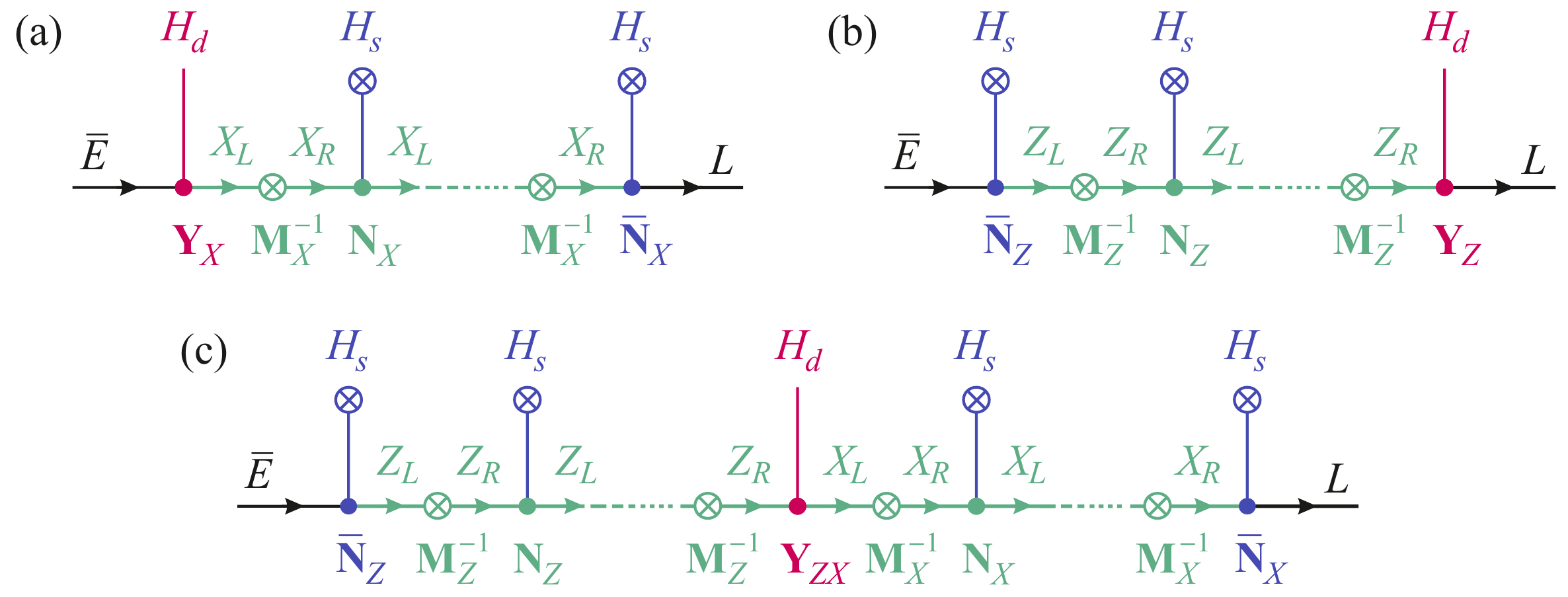}
\caption{The effective contributions to the leptonic Yukawa couplings generated by the tree-level exchanges of (a) a flavor-triplet, weak doublet of vector leptons $X_{L,R}$ or (b) a flavor-triplet, weak singlet of vector leptons $Z_{L,R}$. When both these states are present, the two contributions (a) and (b) are accompanied by their simultaneous exchanges (c).}%
\label{Fig2}
\end{figure}

When the vector leptons are heavy, they can be integrated out by solving their
equations of motion%
\begin{align}
\frac{\delta\mathcal{L}_{\xi}}{\delta X_{L}}  &  =\bar{X}_{R}(\mathbf{M}%
_{X}+\mathbf{N}_{X}H_{s})+\bar{E}\mathbf{Y}_{X}H_{d}=0\;\;,\\
\frac{\delta\mathcal{L}_{\xi}}{\delta X_{R}}  &  =\bar{X}_{L}(\mathbf{M}%
_{X}+\mathbf{N}_{X}^{\dagger}H_{s})+\bar{L}\mathbf{\bar{N}}_{X}^{\dagger}%
H_{s}=0\;.
\end{align}
Plugging this back into the Lagrangian, we get a contribution to the leptonic
Yukawa interaction (see Fig.~\ref{Fig2}a)%
\begin{equation}
\mathbf{Y}_{e}^{eff}=\mathbf{Y}_{e}-\mathbf{Y}_{X}\frac{1}{\mathbf{M}%
_{X}+\mathbf{N}_{X}H_{s}}\mathbf{\bar{N}}_{X}H_{s}=\gamma\mathbf{Y}_{d}%
\cdot\frac{\mathbf{1}}{\mathbf{1}+(v_{s}/M_{X})(\alpha\mathbf{Y}_{u}^{\dagger
}\mathbf{Y}_{u}+\beta\mathbf{Y}_{d}^{\dagger}\mathbf{Y}_{d})}\;, \label{OnlyX}%
\end{equation}
with $v_{s}$ the scalar singlet vacuum expectation value. Provided
$v_{s}/M_{X}\gg1$, this precisely reproduces the geometric sum discussed in
the previous section. Importantly, no resummation was involved: the $X_{L,R}$
mass terms and their interactions with $H_{s}$ were integrated exactly. If
these terms were treated perturbatively, one would recover a geometric MFV
series. So, in this case, the issue of the convergence of the MFV series is
really similar to that of the usual QFT perturbative series.

\begin{figure}[t]
\centering     \includegraphics[width=0.95\textwidth]{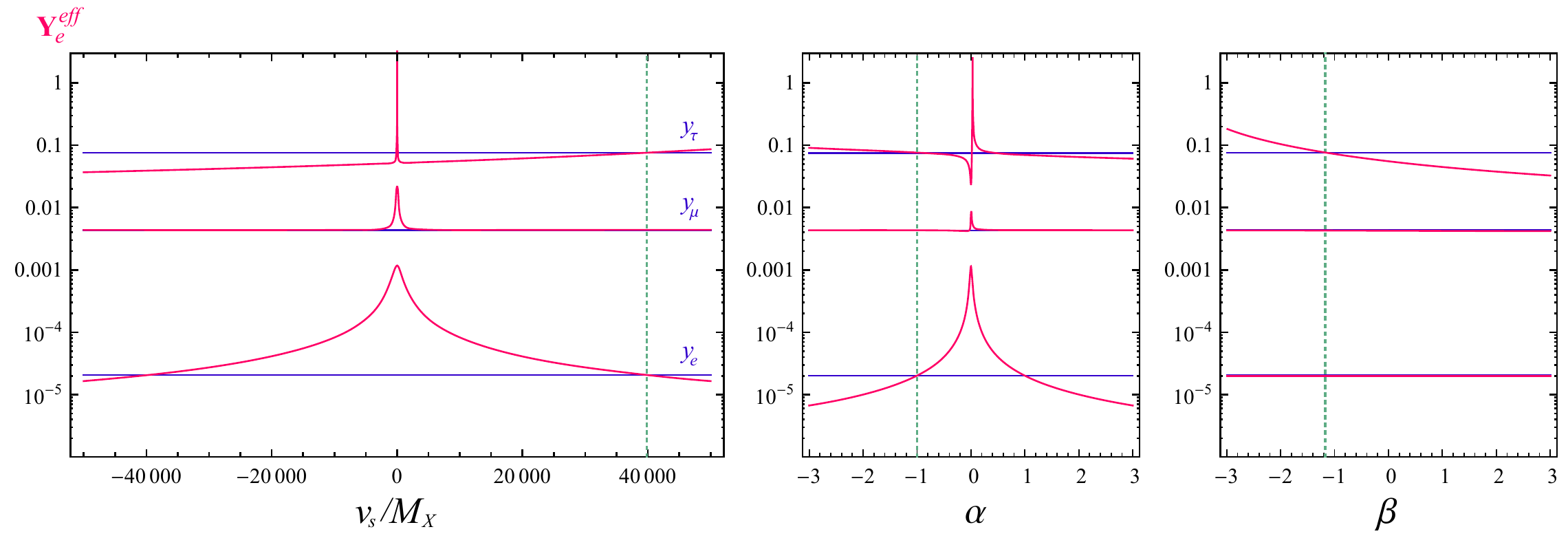}
\caption{Evolution of the singular values of $\mathbf{Y}_{e}^{eff}$ as given in Eq.~\ref{OnlyX}, i.e., induced by tree-level exchanges of the vector leptons $X_{L,R}$. The green dashed line indicates the point at which the singular values of $\mathbf{Y}_{e}^{eff}$ coincide with the observed lepton Yukawa couplings, and corresponds to the values quoted in Eq.~\ref{InGUT1}. Compared to Fig.~\ref{Fig1}, it now sits at a perfectly regular point. The singular values are not fine-tuned since they do not change much as the $\mathbf{Y}_{e}^{eff}$ parameters vary.}
\label{Fig3}
\end{figure}

Numerically, we fix $\alpha$ and solve for the remaining parameters $\gamma$,
$\beta$ and $v$ so that the three singular values $\mathbf{Y}_{e}^{eff}$
reproduce the observed lepton masses. For the MSSM at the GUT
scale\footnote{The values of the Yukawa couplings at the GUT scale quoted in
Ref.~\cite{Bora} used here should only be considered illustrative, since they
do not take into account the presence of the vector leptons at some
intermediate scale.} with $\tan\beta=10$, we find%
\begin{equation}
\gamma=22\;,\;\;\alpha\equiv-1\;,\;\;\beta=-1.2\;,\;\;\frac{v_{s}}{M_{X}%
}=4\times10^{4}\;. \label{InGUT1}%
\end{equation}
The expansion coefficients are very reasonable when the ratio $v_{s}/M_{X}$ is
large. Importantly, the value of $v_{s}$ is totally decorrelated from that of
$\mathbf{Y}_{u}$ or $\mathbf{Y}_{d}$. As shown in Fig.~\ref{Fig3}, the evolution of the
singular values of $\mathbf{Y}_{e}^{eff}$ as $v_{s}$ is varied is rather
smooth over a large range (keep in mind though that the scale of the plot is
logarithmic). The same is true when varying $\alpha$ or $\beta$, ensuring this
solution is free of any fine-tuning.

The lepton mixing matrices at the best-fit point are%
\begin{equation}
|g_{E}|=\left(
\begin{array}
[c]{ccc}%
0.96 & 0.26 & 0.0040\\
0.24 & 0.88 & 0.42\\
0.11 & 0.41 & 0.91
\end{array}
\right)  \;,\;\;|g_{L}|=\left(
\begin{array}
[c]{ccc}%
0.0089 & 0.98 & 0.17\\
0.038 & 0.17 & 0.98\\
1.00 & 0.015 & 0.0036
\end{array}
\right)  \;. \label{LTwist1}%
\end{equation}
Compared to the mixing matrices obtained using the polynomial expansion,
Eq.~(\ref{TwistPoly}), the same twist of the left leptons happens while the
mixing angles are a bit larger (smaller) in the right (left) sector.

\begin{figure}[t]
\centering     \includegraphics[width=0.5\textwidth]{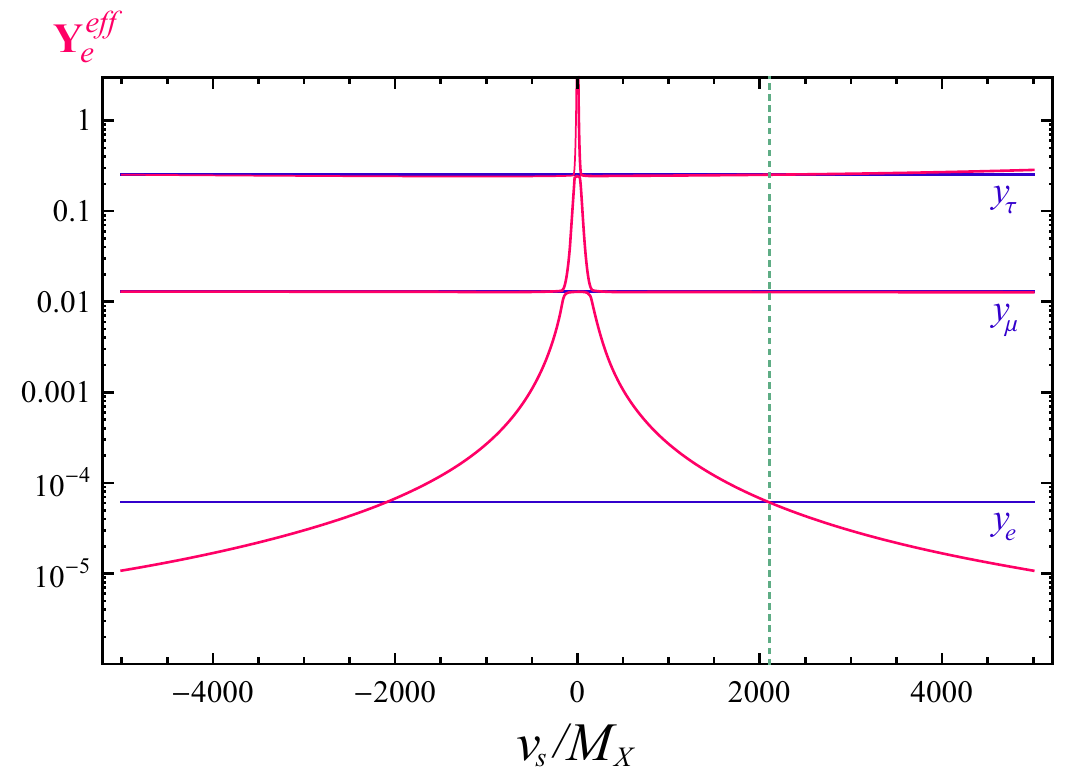}
\caption{Evolution of the singular values of $\mathbf{Y}_{e}^{eff}$ as given in Eq.~\ref{OnlyXZ}, i.e., induced by tree-level exchanges of both the vector leptons $X_{L,R}$ and $Z_{L,R}$. The green dashed line indicates the point at which the singular values of $\mathbf{Y}_{e}^{eff}$ coincide with the observed lepton Yukawa couplings, and corresponds to the values quoted in Eq.~\ref{XZvals}. Compared to Fig.~\ref{Fig1}, it now sits at a perfectly regular point, and compared to Fig.~\ref{Fig3}, even larger departures from that point have a negligible impact on the two largest singular values. The evolution as the other parameters are varied is similar as in Fig.~\ref{Fig3}.}%
\label{Fig4}
\end{figure}

This setting can be generalized in many ways. One interesting extension is to
introduce vector-like partners for both the lepton singlet and doublet. So, we
add a flavor-triplet of vector leptons $Z_{L,R}$ transforming as the
right-handed lepton singlet:%
\begin{align}
\mathcal{L}_{X_{L,R},Z_{L,R}}  &  =\mathcal{L}_{X_{L,R}}+\bar{Z}_{L}(i \slashed{D})Z_{L}+\bar{Z}_{R}(i \slashed{D})Z_{R}+\bar{Z}_{L}\mathbf{M}_{Z}Z_{R}+\bar{Z}_{R}\mathbf{M}_{Z}Z_{L}\frac{{}}{{}}\nonumber\\
&  +\left(  \bar{Z}_{R}\mathbf{N}_{Z}Z_{L}H_{s}+\bar{E}\mathbf{\bar{N}}%
_{Z}Z_{L}H_{s}+\bar{Z}_{R}\mathbf{Y}_{Z}LH_{d}+h.c.\frac{{}}{{}}\right)
\nonumber\\
&  +\left(  \bar{Z}_{R}\mathbf{Y}_{ZX}X_{L}H_{d}+\bar{X}_{R}\mathbf{Y}%
_{XZ}Z_{L}H_{d}+h.c.\frac{{}}{{}}\right)  \;. \label{LagrXZ}%
\end{align}
Choosing now the flavor symmetry as%
\begin{equation}
G_{F}^{\prime}=U\left(  3\right)  _{Q=L=X_{L,R}}\otimes U\left(  3\right)
_{U}\otimes U\left(  3\right)  _{D=E=Z_{L,R}}\;,
\end{equation}
the MFV assumptions become%
\begin{equation}
\left\{
\begin{array}
[c]{c}%
\mathbf{M}_{X,Z}=M_{X,Z}\mathbf{1\;,}\\
\mathbf{Y}_{e}=\mathbf{Y}_{X}=\mathbf{Y}_{Z}=\mathbf{Y}_{XZ}=\mathbf{Y}%
_{ZX}=\gamma\mathbf{Y}_{d}\;,\\
\mathbf{N}_{X}=\mathbf{\bar{N}}_{X}=\alpha\mathbf{Y}_{u}^{\dagger}%
\mathbf{Y}_{u}+\beta\mathbf{Y}_{d}^{\dagger}\mathbf{Y}_{d}\;,\\
\mathbf{N}_{Z}=\mathbf{\bar{N}}_{Z}=\varepsilon\mathbf{Y}_{d}\mathbf{Y}%
_{d}^{\dagger}\;.
\end{array}
\right.  \label{CondXZ}%
\end{equation}
The equations of motion for the four families of heavy leptons $X_{L,R}$ and $Z_{L,R}$ are
coupled because of the mixing term $\mathbf{Y}_{XZ}$ and $\mathbf{Y}_{ZX}$ but
can be solved to first order in $H_{d}$ (see Fig.~\ref{Fig2}):%
\begin{align}
\mathbf{Y}_{e}^{eff}  &  =\mathbf{Y}_{e}-\mathbf{Y}_{X}\frac{1}{\mathbf{M}%
_{X}+\mathbf{N}_{X}H_{s}}\mathbf{\bar{N}}_{X}H_{s}-H_{s}\mathbf{\bar{N}}%
_{Z}\frac{1}{\mathbf{M}_{Z}+\mathbf{N}_{Z}H_{s}}\mathbf{Y}_{Z}\nonumber\\
&  \;\;\;\;\;\;\;\;\;\;\;\;\;\;\;\;\;\;\;\;+H_{s}\mathbf{\bar{N}}_{Z}%
\frac{1}{\mathbf{M}_{Z}+\mathbf{N}_{Z}H_{s}}\mathbf{Y}_{ZX}\frac{1}%
{\mathbf{M}_{X}+\mathbf{N}_{X}H_{s}}\mathbf{\bar{N}}_{X}H_{s}%
\nonumber\label{OnlyXZ}\\
&  =\frac{1}{\mathbf{1}+(v_{s}/M_{Z})(\varepsilon\mathbf{Y}_{d}\mathbf{Y}%
_{d}^{\dagger})}\cdot\gamma\mathbf{Y}_{d}\cdot\frac{1}{\mathbf{1}+(v_{s}%
/M_{X})(\alpha\mathbf{Y}_{u}^{\dagger}\mathbf{Y}_{u}+\beta\mathbf{Y}%
_{d}^{\dagger}\mathbf{Y}_{d})}\;.
\end{align}
By trial and error, we find for example for the MSSM at the GUT scale and
$\tan\beta=30$,%
\begin{equation}
\gamma=81\;,\;\alpha\equiv1\;,\;\;\beta\equiv1\;,\;\;\;\varepsilon
\equiv-1\;,\;\;\;\frac{v_{s}}{M_{X}}=2.1\times10^{3}\;,\;\;\frac{M_{X}}{M_{Z}%
}\equiv7\;,\label{XZvals}
\end{equation}
or%
\begin{equation}
\gamma=83\;,\;\alpha\equiv1\;,\;\;\beta\equiv0\;,\;\;\;\varepsilon
\equiv-1\;,\;\;\;\frac{v_{s}}{M_{X}}=1.1\times10^{4}\;,\;\;\frac{M_{X}}{M_{Z}%
}\equiv0.3\;.
\end{equation}
Infinitely many other solutions exists, some may give slightly lower
$v_{s}/M_{X,Z}$, but none should decrease it dramatically. Concentrating on
the first solution, we show in Fig.~\ref{Fig4} the behavior as $v_{s}$ varies holding the
other parameters fixed. It is evidently free of any fine-tuning, and even more
stable than before. Further, this solution has one very interesting feature.
Once the right-handed sector becomes tuned by a geometric expansion, both
species of leptons end up similarly twisted:%
\begin{equation}
|g_{E}|=\left(
\begin{array}
[c]{ccc}%
0.015 & 1.00 & 0.0019\\
0.0032 & 0.0018 & 1.00\\
1.00 & 0.015 & 0.003
\end{array}
\right)  \;,\;\;|g_{L}|=\left(
\begin{array}
[c]{ccc}%
0.0069 & 1.00 & 0.019\\
0.032 & 0.020 & 1.00\\
1.00 & 0.0075 & 0.031
\end{array}
\right)  \;. \label{LTwist2}%
\end{equation}
The mixing angles are also greatly reduced. This means that in this scenario,
the true identity of the electron is completely altered: it is mostly the
third-generation gauge state.

These constructions are not meant to be full-fledged models. Rather, they are
designed to illustrate the main mechanism by which the lepton and quark flavor
structures could be related in a natural way. The most salient features are

\begin{itemize}
\item The value for $\gamma$ is often found a bit too large if one has in mind
GUT settings where the boundary conditions set e.g. $\mathbf{Y}_{e}%
^{T}=\mathbf{Y}_{d}$. Still, the situation is different in GUT since both
$\mathbf{Y}_{e}$ and $\mathbf{Y}_{d}$ have to be generated simultaneously out
of other Yukawa couplings. This will be discussed in Section~\ref{SecGUT}. Also, this
value of $\gamma$ depends quite crucially on the value of $v_{s}$, and on the
MFV conditions Eq.~(\ref{CondX}) or~(\ref{CondXZ}).

\item The effective contribution to $\mathbf{Y}_{e}$ decouples when either
$v_{s}\rightarrow0$ or $M_{X,Z}\rightarrow\infty$, but is non-decoupling in
the $v_{s}>M_{X,Z}$ region relevant for the geometric behavior.

\item The scales $v_{s}$ and $M_{X,Z}$ are free since only their ratio plays a
role. Further, any rescaling of the expansion coefficients in $\mathbf{N}%
_{X,Z}$ and $\mathbf{\bar{N}}_{X,Z}$ can be compensated by a change in
$v_{s}/M_{X,Z}$. In particular, one could imagine that in some more complete
model, $\mathbf{N}_{X,Z}$ and $\mathbf{\bar{N}}_{X,Z}$ are radiatively
induced. This would naturally explain the specific form of their expansions in
Eq.~(\ref{CondX}) or~(\ref{CondXZ}), at the cost of further increasing
$v_{s}/M_{X,Z}$.

\item Such large $v_{s}/M_{X,Z}$ ratios imply that $v_{s}\gg v_{u,d}$ since
$M_{X,Z}$ should be above the EW scale for all the vector leptons to be
integrated out. If not protected by some symmetry, this large hierarchy could
require delicate fine-tunings in the scalar sector, in case $H_{s}$ couples to
$H_{u}$ and/or $H_{d}$. Note though that in a supersymmetric setting, the only
allowed superpotential term would be $H_{s}H_{u}H_{d}$, which breaks the
Peccei-Quinn symmetry~\cite{PecceiQ77} if the couplings in Eq.~(\ref{LagrXZ})
do not. Connection with invisible axion models~\cite{Axions}, where large
hierarchies are also present between symmetry breaking scales, could offer
interesting perspective, which we leave for future works.

\item The MFV conditions Eq.~(\ref{CondX}) or~(\ref{CondXZ}) could be thought
of as boundary conditions, in a way similar to the mSUGRA pattern for
supersymmetry breaking terms. But, looking at Eq.~(\ref{OnlyX}) or
Eq.~(\ref{CondXZ}), it is clear that even small deviations from these
conditions would spoil the geometric behavior for the MFV series.
\end{itemize}

Let us analyze this last point in a bit more detail. Indeed, it would not be
very convincing to trade the fine-tunings in the coefficients in
Eq.~(\ref{CoefPoly}) for a fine-tuning in the boundary conditions. Ultimately,
the mechanism at the origin of these conditions should be accompanied by a
symmetry able to stabilize them, at least partially. Looking back at
Eq.~(\ref{LagrXZ}), we can already glean some hints of how this could arise.
If we combine together the nine weak doublet lepton fields into $\Phi_{L}%
^{T}=(L,X_{L},X_{R})$ and the nine weak singlet lepton fields into $\Phi
_{R}^{T}=(E,Z_{R},Z_{L})$, then
\begin{equation}
\mathcal{L}_{X_{L,R},Z_{L,R},E,L}=\bar{\Phi}_{L}(i \slashed{D}+\mathbf{P}_{X})\Phi_{L}+\bar{\Phi}_{R}(i \slashed{D} +\mathbf{P}_{Z})\Phi_{R}+(\bar{\Phi}_{R}\mathbf{Y}\Phi_{L}H_{d}+h.c.)\;,
\label{U91}%
\end{equation}
actually exhibits a $U(9)_{L}\otimes U(9)_{R}$ flavor symmetry broken only by
the flavor structures:%
\begin{equation}
\mathbf{P}_{A}=\left(
\begin{array}
[c]{ccc}%
0 & 0 & 0\\
0 & 0 & \mathbf{M}_{A}\\
0 & \mathbf{M}_{A} & 0
\end{array}
\right)  +\left(
\begin{array}
[c]{ccc}%
0 & 0 & \mathbf{\bar{N}}_{A}\\
0 & 0 & \mathbf{N}_{A}\\
\mathbf{\bar{N}}_{A} & \mathbf{N}_{A} & 0
\end{array}
\right)  H_{s},\;\;\mathbf{Y}=\left(
\begin{array}
[c]{ccc}%
\mathbf{Y}_{e} & \mathbf{Y}_{X} & 0\\
\mathbf{Y}_{Z} & \mathbf{Y}_{ZX} & 0\\
0 & 0 & \mathbf{Y}_{XZ}%
\end{array}
\right)  \;. \label{U92}%
\end{equation}
Most vanishing entries are due to chirality, the rest to the gauge symmetries.
Consider then the $\mathbf{M}_{X,Z}=0$ limit. The MFV conditions
Eq.~(\ref{CondX}) or~(\ref{CondXZ}) emerge as the only one invariant under
$(X_{L}\leftrightarrow L)$ and $(Z_{R}\leftrightarrow E)$. Imagine thus that
$H_{s}$ first gets its VEV at the scale $v_{s}$, while this discrete symmetry
is broken spontaneously at the much lower scale $M_{X,Z}$. The deviations with
respect to the conditions Eq.~(\ref{CondX}) or~(\ref{CondXZ}) would end up
tiny, at most of the order of $M_{X,Z}/v_{s}$, and would not completely alter
the $v_{s}$ scaling of $\mathbf{Y}_{e}^{eff}$. Actually, such corrections may
even be welcome to reduce the numerical value of $\gamma$ or the ratio
$v_{s}/M_{X,Z}$.

\section{Scenario 2: Supersymmetry and light stops}

Supersymmetry is one the most studied extension to the SM. Besides its
intrinsic mathematical appeal, it is able to solve, or at least lessen,
several puzzles of the SM, and most notably the issue of the stability of the
electroweak scale. At the same time, low-scale supersymmetry is expected to
influence various flavor physics observables, and its many new states are
within range of direct production at the LHC. The absence of any signal up to
now puts strong constraints on viable supersymmetric scenarios. Our goal in this
section is to analyze in which respect the relationship discovered between
$\mathbf{Y}_{e}$ and $\mathbf{Y}_{u,d}$ could help.

\subsection{Squark mass matrices with geometric expansions}

Direct searches for supersymmetric particles at colliders are particularly
sensitive to first-generation squarks, simply because of the presence of many
such quarks in the initial state. The current bounds are typically well above
1 TeV, depending on the assumptions on the masses of the other
sparticles~\cite{SUSYAltas,SUSYCMS}. On the contrary, for third generation
squarks, the bounds are still below the TeV. In this context, Natural
SUSY-like scenarios~\cite{NSUSY} where third generation squarks are much
lighter than the others offer interesting settings. What we now want to show
is that settings where all MFV expansions are geometric actually generate such patterns.

Consider the following geometric MFV parametrization for the squark
soft-breaking terms%
\begin{equation}
\mathbf{m}_{Q,U,D}^{2}=m_{0}^{2}\mathbf{X}_{Q,U,D}\;\;,\;\;\;\mathbf{A}%
_{u,d}=A_{0}\mathbf{X}_{U,D}\cdot\mathbf{Y}_{u,d}\cdot\mathbf{X}_{Q}\;,
\label{SoftQ}%
\end{equation}
with
\begin{subequations}
\label{SoftXDUD}%
\begin{align}
\mathbf{X}_{Q}  &  =\frac{1}{\mathbf{1}-\eta_{q}(\alpha_{1}^{q}\mathbf{Y}%
_{u}^{\dagger}\mathbf{Y}_{u}+\alpha_{2}^{q}\mathbf{Y}_{d}^{\dagger}%
\mathbf{Y}_{d}+...)}\;,\\
\mathbf{X}_{U}  &  =\frac{1}{\mathbf{1}-\eta_{u}(\alpha_{1}^{u}\mathbf{Y}%
_{u}\mathbf{Y}_{u}^{\dagger}+\alpha_{2}^{u}\mathbf{Y}_{u}\mathbf{Y}%
_{d}^{\dagger}\mathbf{Y}_{d}\mathbf{Y}_{u}^{\dagger}+...)}\;,\\
\mathbf{X}_{D}  &  =\frac{1}{\mathbf{1}-\eta_{d}(\alpha_{1}^{d}\mathbf{Y}%
_{d}\mathbf{Y}_{d}^{\dagger}+\alpha_{2}^{d}\mathbf{Y}_{d}\mathbf{Y}%
_{u}^{\dagger}\mathbf{Y}_{u}\mathbf{Y}_{d}^{\dagger}+...)}\;,
\end{align}
where $\alpha_{i}^{q,u,d}$ are $\mathcal{O}(1)$ parameters, $m_{0}$, $A_{0}$
the SUSY-breaking scale parameters, which we set at $m_{0}=10$ TeV, $A_{0}=-1$
TeV, and we assume $\mathbf{X}_{Q,U,D}$ are the same when entering squared
squark masses or trilinear terms for simplicity

Though we will not attempt at constructing a fully dynamical model, it is
tempting to think of these $\mathbf{X}_{Q}$, $\mathbf{X}_{U}$, and
$\mathbf{X}_{D}$ factors as arising from the exchange of new states whose
propagators transform like $SU(3)_{Q}$, $SU(3)_{U}$ or $SU(3)_{D}$ octets,
respectively. The corresponding coefficients $\eta_{q,u,d}=v_{s}/M_{X_{Q,U,D}%
}$, with $M_{X_{Q,U,D}}$ the mass of these octets and $v_{s}$ the VEV of some
singlet Higgs bosons, can in principle be large. These $\mathbf{X}_{Q,U,D}$
factor then match those studied in the previous section, with for example%
\end{subequations}
\begin{equation}
\mathbf{X}_{U}\approx\mathbf{1}-\frac{1}{\langle\mathbf{Y}_{u}^{\dagger
}\mathbf{Y}_{u}\rangle}\mathbf{Y}_{u}\mathbf{Y}_{u}^{\dagger}\;.
\end{equation}
Note that it may make more sense to think of these new states as scalars, in
which case $\mathbf{X}_{Q,U,D}^{2}$ propagator factors would appear in
Eq.~(\ref{SoftQ}). Numerically, this would not change much the boundary
conditions for the squark soft-breaking terms since the strict
third-generation dominance approximation $(\mathbf{Y}_{i}^{\dagger}%
\mathbf{Y}_{i})^{2}\rightarrow\langle\mathbf{Y}_{i}^{\dagger}\mathbf{Y}%
_{i}\rangle\mathbf{Y}_{i}^{\dagger}\mathbf{Y}_{i}$ implies for example
\begin{equation}
\mathbf{X}_{U}^{2}\approx\left(  \mathbf{1}-\frac{1}{\langle\mathbf{Y}%
_{u}^{\dagger}\mathbf{Y}_{u}\rangle}\mathbf{Y}_{u}\mathbf{Y}_{u}^{\dagger
}\right)  ^{2}\approx\mathbf{1}-\frac{1}{\langle\mathbf{Y}_{u}^{\dagger
}\mathbf{Y}_{u}\rangle}\mathbf{Y}_{u}\mathbf{Y}_{u}^{\dagger}\;.
\end{equation}
For simplicity, we thus stick to the linear expansions in Eq.~(\ref{SoftQ}).

In the large $\eta_{q,u,d}$ limit, this setting actually matches that studied
in Ref.~\cite{NatSUSY} from a purely phenomenological perspective. There, the
large $\eta_{q,u,d}$ limit of the expansions in Eq.~(\ref{SoftQ}) were imposed
at the GUT scale and evolved down to the TeV scale. Let us summarize the main results:

\begin{itemize}
\item To end up with only the $\tilde{t}_{L}$ and $\tilde{b}_{L}$ as light
states, one should set $\eta_{u,d}=0$, $\eta_{q}\gg1$. However, the RG
evolution necessarily drives the small $\mathbf{m}_{Q}^{2}[M_{GUT}]^{33}$
towards negative values. This results in an unacceptable color-breaking
minimum. To prevent this, either one should impose Eq.~(\ref{SoftQ}) at a much
lower scale, or $\mathbf{m}_{U}^{2}[M_{GUT}]^{33}$ must also be small. In this
latter case, setting $\eta_{u}=\eta_{q}\gg1$, the three squark states
$\tilde{t}_{L,R}$ and $\tilde{b}_{L}$ end up much lighter than the other
squarks, whose masses remain very close to $m_{0}$.

\item Except at very large $\tan\beta$, the impact of $\mathbf{X}_{D}$ is
always negligible and $\tilde{b}_{R}$ remains quasi-degenerate with the first-
and second-generation squarks.

\item The RG evolution of the trilinear terms $\mathbf{A}_{U,D}$ wipes out the
effect of the $\mathbf{X}_{Q,U,D}$ factors. In other words, at the low scale,
the trilinear terms obtained either from $\mathbf{A}_{U,D}[M_{GUT}%
]=A_{0}\mathbf{X}_{U,D}\cdot\mathbf{Y}_{u,d}\cdot\mathbf{X}_{Q}$ or simply
from $\mathbf{A}_{U,D}[M_{GUT}]=A_{0}\mathbf{Y}_{u.d}$ are very similar, and
so are the resulting squark mass spectra.

\item Because $\langle\mathbf{Y}_{u}^{\dagger}\mathbf{Y}_{u}\rangle$ and
$\langle\mathbf{Y}_{d}^{\dagger}\mathbf{Y}_{d}\rangle$ are at most
$\mathcal{O}(1)$, these expansions satisfy the usual MFV naturality
requirement. As a result, supersymmetric contributions to flavor transitions
remain tuned by the CKM matrix, and the constraints from flavor observables
are satisfied even with rather light sparticles.
\end{itemize}

\subsection{Untwisted slepton mass matrices and $\mu\rightarrow e\gamma$}

To express the lepton Yukawa coupling in terms of those of the quarks, the
flavor symmetry was reduced to $G_{F}^{\prime}=U\left(  3\right)
_{Q=L}\otimes U\left(  3\right)  _{U}\otimes U\left(  3\right)  _{D=E}$. In a
supersymmetric setting, $G_{F}^{\prime}$ also allows for the slepton
soft-breaking terms to be expressed in terms of $\mathbf{Y}_{u,d}$.
Altogether, the lepton and slepton flavor-breaking sector becomes%
\begin{equation}
\mathbf{Y}_{e}=\gamma\mathbf{X}_{D}\cdot\mathbf{Y}_{d}\cdot\mathbf{X}_{Q}\;,
\end{equation}
and%
\begin{equation}
\mathbf{m}_{L}^{2}=m_{0}^{2}\mathbf{X}_{Q}\;,\;\;\mathbf{m}_{E}^{2}=m_{0}%
^{2}\mathbf{X}_{D}\;,\;\;\mathbf{A}_{e}=\gamma A_{0}\;\mathbf{X}_{D}%
\cdot\mathbf{Y}_{d}\cdot\mathbf{X}_{Q}\;. \label{mLGeo}%
\end{equation}
For simplicity, we assume universal expansions in each $SU(3)$ sectors, i.e.,
all $\mathbf{X}_{Q}$ factors are identical, and so are all the $\mathbf{X}%
_{E}$.

Once these conditions are set, the freedom to rotate the (s)lepton doublet and
singlet is recovered since the MSSM exhibit a $G_{F}$ symmetry in its gauge
sector. Thus, $\mathbf{Y}_{e}$ can be diagonalized through%
\begin{equation}
L\rightarrow g_{L}L,E\rightarrow g_{E}E,\;\;\mathbf{Y}_{e}\rightarrow
(\mathbf{Y}_{e})^{phys}=g_{E}\mathbf{Y}_{e}g_{L}^{\dagger}\;,
\end{equation}
with $(\mathbf{Y}_{e})^{phys}=\operatorname{diag}(y_{e},y_{\mu},y_{\tau})$.
This same rotation has to be performed on the slepton partners, so that in the
lepton physical basis,%
\begin{equation}
(\mathbf{m}_{L}^{2})^{phys}=m_{0}^{2}g_{L}\mathbf{X}_{Q}g_{L}^{\dagger
}\;,\;\;(\mathbf{m}_{E}^{2})^{phys}=m_{0}^{2}g_{E}\mathbf{X}_{D}g_{E}%
^{\dagger}\;,\;\;(\mathbf{A}_{e})^{phys}=A_{0}\mathbf{Y}_{e}\;.
\end{equation}
This action of the mixing matrices $g_{L}$ and $g_{E}$ has two
particularities. First, neither $(\mathbf{m}_{L}^{2})^{phys}$ nor
$(\mathbf{m}_{E}^{2})^{phys}$ are diagonal in general, since the matrices
$g_{E}$ and $g_{L}$ come from the SVD of $\mathbf{Y}_{e}$. Their off-diagonal
entries are of the order of CKM entries, since they are generated by the
mismatch between $\mathbf{Y}_{u}$ and $\mathbf{Y}_{d}$ entering in
$\mathbf{X}_{Q,D}$. Second, even if $g_{L}$ and/or $g_{E}$ can twist the
leptons, as in Eq.~(\ref{LTwist1}) or Eq.~(\ref{LTwist2}), this same twist is
then enforced on their supersymmetric partners. For example, if only
$\mathbf{X}_{Q}$ is present, $(e_{L}^{-})^{phys}$ is essentially the
$(\tau_{L})^{gauge}$ state, then $(\tilde{e}_{L})^{phys}$ is essentially the
$(\tilde{\tau}_{L})$ gauge state. Further, given that
\begin{equation}
1\approx(\mathbf{m}_{L}^{2})^{11}\approx(\mathbf{m}_{L}^{2})^{22}%
\gg(\mathbf{m}_{L}^{2})^{33}\approx0\;,
\end{equation}
the $(\tilde{e}_{L})^{phys}$ and $(\tilde{\nu}_{L})^{phys}$ states are much
lighter than the other sleptons.

Non-vanishing off-diagonal entries in $(\mathbf{m}_{L}^{2})^{phys}$ together with rather light first-generation sleptons immediately raise the question of lepton flavor violating (LFV) observables. A process like $\mu\rightarrow e\gamma$ can be induced by neutralino and chargino loops, with a branching ratio scaling like~\cite{CERNleptons}%
\begin{equation}
\mathcal{B}(\mu\rightarrow e\gamma)\approx10^{-5}\frac{M_{W}^{4}}%
{M_{\tilde{\ell}}^{8}}\tan\beta\left|  (\mathbf{m}_{L}^{2})_{12}%
^{phys}\right|  ^{2}F_{SUSY}\;,
\end{equation}
where $M_{\tilde{\ell}}$ is the typical slepton mass, which we take as the
geometric average of the involved sleptons, $M_{\tilde{\ell}}^{2}=M_{\tilde
{e}}M_{\tilde{\mu}}$, and $F_{SUSY}$ is an $\mathcal{O}(1)$ function of the
sparticle masses. The question is then whether the current bound
$\mathcal{B}(\mu\rightarrow e\gamma)^{\exp}<5.7\times10^{-13}$~\cite{MEG} is satisfied.

\begin{figure}[t]
\centering     \includegraphics[width=0.5\textwidth]{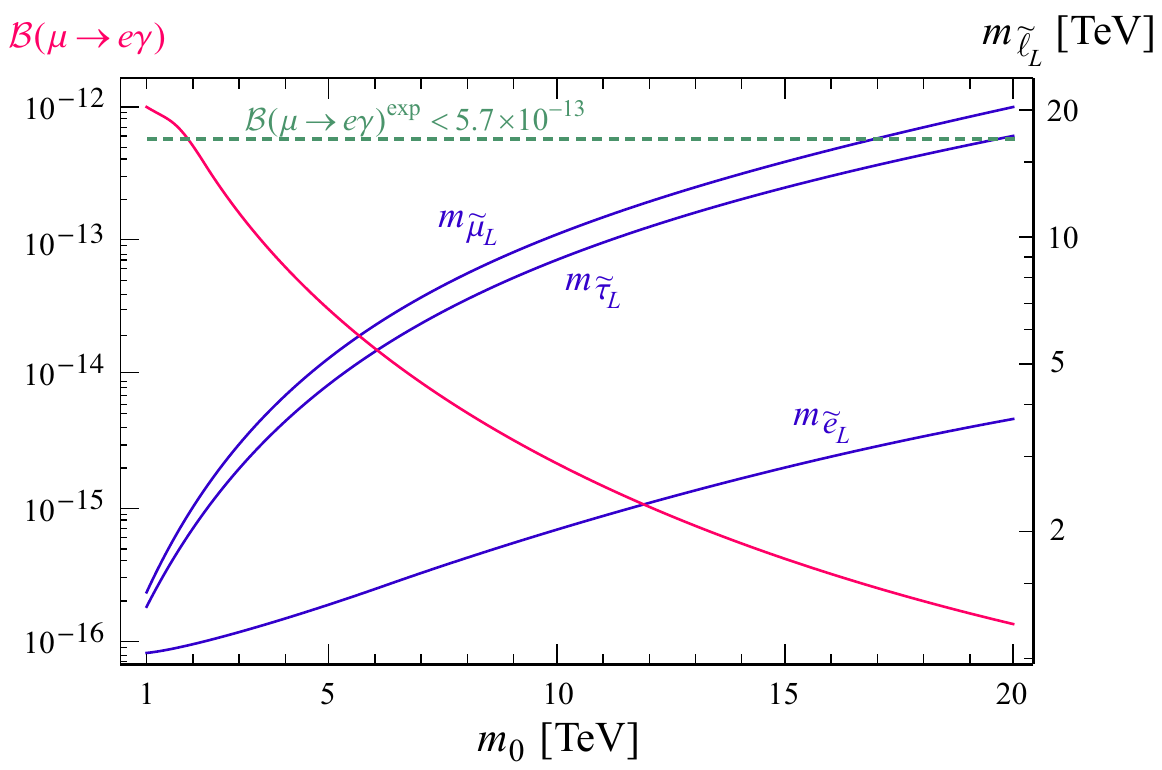}
\caption{Left slepton masses and $\mathcal{B}(\mu\rightarrow e\gamma)$ as a function of the SUSY breaking scale $m_0$, with $(\mathbf{m}_{L}^{2}[1\,\text{TeV}])^{phys}$ given in Eq.~(\ref{SleptPhys}).}%
\label{Fig5}
\end{figure}

To illustrate that this is indeed the case, let us consider a specific
realization. We set the boundary conditions at the GUT scale, and perform the
evolution at NLO. For simplicity, we introduce only $\mathbf{X}_{Q}$ and not
$\mathbf{X}_{D}$. The inputs at the GUT scale are slightly different than for
Eq.~(\ref{InGUT1}), because of the specific MSSM parameters chosen
here\footnote{The MSSM parameters are fixed assuming a CMSSM-like setting,
with $A_{0}=-1\,$TeV, $m_{H_{u}}^{2}=m_{H_{d}}^{2}=2\times10^{6}\,$GeV$^{2}$,
$M_{1/2}=1.5$ $\,$TeV and $\tan\beta=10$. The parameter $m_{0}$, setting the
scale of both squark and slepton soft-terms, is allowed to vary. At the GUT
scale, we set $\mathbf{m}_{Q,U,D,E}^{2}=m_{0}^{2}\mathbf{1}$, $A_{u,d,e}%
=A_{0}\mathbf{Y}_{u,d,e}$ and $\mathbf{m}_{L}^{2}$ as in Eq.~(\ref{mLGeo}).},
and we take%
\begin{equation}
\gamma=19\;,\;\;\alpha_{1}^{q}\equiv1\;,\;\;\alpha_{2}^{q}=2.6\;,\;\;\eta
_{q}=2.1\times10^{4}\;. \label{SleptRG}%
\end{equation}
Once $\mathbf{Y}_{e}=\gamma\mathbf{Y}_{d}\cdot\mathbf{X}_{Q}$ is fixed,
$g_{L}$ and $\mathbf{m}_{L}^{2}=m_{0}^{2}\mathbf{X}_{Q}$ can be computed
directly:%
\begin{equation}
|g_{L}|=\left(
\begin{array}
[c]{ccc}%
0.0098 & 0.99 & 0.17\\
0.041 & 0.17 & 0.98\\
1.00 & 0.017 & 0.045
\end{array}
\right)  ,\;\;|\mathbf{m}_{L}^{2}[M_{GUT}]|=m_{0}^{2}\left(
\begin{array}
[c]{ccc}%
0.99 & 0.045 & 0.0066\\
0.045 & 0.74 & 0.029\\
0.0066 & 0.029 & 0.0014
\end{array}
\right)  \;,
\end{equation}
so%
\begin{equation}
|(\mathbf{m}_{L}^{2}[M_{GUT}])^{phys}|=m_{0}^{2}\left(
\begin{array}
[c]{ccc}%
0.00049 & 0.018 & 0.0015\\
0.018 & 1.00 & 0.0029\\
0.0015 & 0.0029 & 0.74
\end{array}
\right)  \;.
\end{equation}
Notice how acting with $g_{L}$ reorders the entries of $\mathbf{m}_{L}^{2}$.
Evolving down, only the diagonal entries are significantly affected since
$\mathbf{Y}_{e}$ is diagonal in the physical basis at all scale. For example,
with $m_{0}=10$ TeV, we find at the low-scale,%
\begin{equation}
|(\mathbf{m}_{L}^{2}[1\,\text{TeV}])^{phys}|=m_{0}^{2}\left(
\begin{array}
[c]{ccc}%
0.042 & 0.018 & 0.0015\\
0.018 & 1.04 & 0.0029\\
0.0015 & 0.0029 & 0.77
\end{array}
\right)  \;. \label{SleptPhys}
\end{equation}
This means that the current bound on $\mathcal{B}(\mu\rightarrow e\gamma)$
translate as a lower bound on $m_{0}$. Coincidentally, as plotted in Fig.~\ref{Fig5},
the current limit does not constrain $m_{0}$ much yet, but any improvement on
$\mathcal{B}(\mu\rightarrow e\gamma)^{\exp}$ would start to push $m_{0}$ well
beyond $1$ TeV. Note, finally, that setting $\mathbf{m}_{L}^{2}=m_{0}%
^{2}\mathbf{X}_{Q}^{2}$ instead of $\mathbf{m}_{L}^{2}=m_{0}^{2}\mathbf{X}%
_{Q}$ does not impact $\mathcal{B}(\mu\rightarrow e\gamma)$ significantly,
since it would mean setting
\begin{equation}
|(\mathbf{m}_{L}^{2}[M_{GUT}])^{phys}|=m_{0}^{2}\left(
\begin{array}
[c]{ccc}%
0.00035 & 0.018 & 0.0052\\
0.018 & 1.01 & 0.016\\
0.0052 & 0.016 & 2.44
\end{array}
\right)  \;.
\end{equation}

In conclusion, the supersymmetric implications of the redundancy of
$\mathbf{Y}_{e}$ opens the way for sizable LFV processes. Even in a simplified
CMSSM-like setting, the current bounds on these modes start to be competitive
in setting constraints on the viable parameter space. A full analysis,
including non-universal squark mass terms, collider, flavor, and Higgs sector
constraints would be in order at this stage, but this is left for future studies.

\subsection{Effectively holomorphic R-parity violation}

Another path to understand the current absence of supersymmetric signals at
the LHC is to give up R parity. In that case, sparticles would decay, the
lightest supersymmetric particle (LSP) may not be neutral and colorless, and
typical missing energy signatures would disappear. Instead, supersymmetry would
show up in hadronic channels, most notably in the same-sign top quark pair
production. Current bounds from these signatures are below the
TeV~\cite{DurieuxS13,ATLASSSL}.

\subsubsection{The MFV alternative to R-parity}

Once the ad-hoc R-parity is removed, the proton ceased to be stable but MFV
has been shown to suppress its rate down to acceptable
levels~\cite{NikolidakisS07}. Indeed, the MSSM spurion content, $\mathbf{Y}%
_{u,d,e}$, does not permit to construct lepton-number violating ($\Delta
\mathcal{L}$) couplings%
\begin{equation}
\mathcal{W}_{\Delta\mathcal{L}}=\frac{1}{2}\boldsymbol{\lambda}  ^{IJK}%
\tilde{L}^{I}\tilde{L}^{J}\tilde{E}^{K}+\boldsymbol{\lambda}  ^{\prime
IJK}\tilde{L}^{I}\tilde{Q}^{J}\tilde{D}^{K}+\boldsymbol{\mu}  ^{\prime I}%
H_{u}\tilde{L}^{I}\;, \label{susy1}%
\end{equation}
but does allow for baryon number violating ($\Delta\mathcal{B}$) couplings%
\begin{equation}
\mathcal{W}_{\Delta\mathcal{B}}=\frac{1}{2}\boldsymbol{\lambda}
^{\prime\prime IJK}\tilde{U}^{I}\tilde{D}^{J}\tilde{D}^{K}\;, \label{susy2}%
\end{equation}
where $\tilde{L}$, $\tilde{Q}$ are the superfields containing the left
doublets $L$, $Q$ and $\tilde{U}$, $\tilde{D}$, $\tilde{E}$ those containing
the left singlets $U^{\dagger}$, $D^{\dagger}$, $E^{\dagger}$. For example, we
can write%
\begin{equation}
\boldsymbol{\lambda}  ^{\prime\prime IJK}=\lambda\varepsilon^{LMN}%
\mathbf{Y}_{u}^{IL}\mathbf{Y}_{d}^{JM}\mathbf{Y}_{d}^{KN}\;. \label{Holo1}%
\end{equation}
It is only once neutrino acquire a Majorana mass term, then included among the
spurions, that $\Delta\mathcal{L}$ couplings are permitted but they end up
sufficiently tiny to pass all the proton decay bounds.

In the present work, we take as spurions only $\mathbf{Y}_{u,d}$, and reduce
the symmetry group to $G_{F}^{\prime}=U\left(  3\right)  _{Q=L}\otimes
U\left(  3\right)  _{U}\otimes U\left(  3\right)  _{D=E}$, which we now
further reduce to%
\begin{equation}
G_{F}^{\prime}=SU\left(  3\right)  _{Q=L}\otimes SU\left(  3\right)
_{U}\otimes SU\left(  3\right)  _{D=E}\;,
\end{equation}
to allow for $\mathcal{B}$ and/or $\mathcal{L}$ violation. Interestingly,
relating the quark and lepton flavor groups does not open the way for
$\Delta\mathcal{L}$ couplings. It is still impossible to construct them out of
the $\mathbf{Y}_{u,d}$ spurions in a $G_{F}^{\prime}$-symmetric way.
Ultimately, the reason for this is the selection rules imposed by
MFV~\cite{MFVBL}. Because $\mathbf{Y}_{u,d}$ transform according to
fundamental representations, and because $\Delta\mathcal{B}\neq0$ or
$\Delta\mathcal{L}\neq0$ requires some contractions with the antisymmetric
$SU(3)$ invariant tensor, $\mathcal{L}$ and $\mathcal{B}$ are broken in
multiples of three elementary units. Each (s)quarks has $\mathcal{B}=1/3$ and
each (s)lepton has $\mathcal{L}=1$, so the selection rules are $\Delta
\mathcal{B}=3n(1/3)=n$ but $\Delta\mathcal{L}=3n$ for any integer $n$. This is
not compatible with $\mathcal{W}_{\Delta\mathcal{L}}$, which breaks
$\mathcal{L}$ by only one elementary unit.

\subsubsection{Holomorphy beats geometric MFV}

The MFV parametrization in Eq.~(\ref{Holo1}) has the interesting property to
be holomorphic in the spurions~\cite{CsakiGH11}. This means that if these
become true dynamical fields at some scale, this term would be the only one
allowed. Further, this property renders the RG evolution of this coupling
particularly simple, and effectively it acts as a powerful IR
attractor~\cite{BernonS14}.

This reasoning is a bit orthogonal to the philosophy followed here. Since
$\mathbf{Y}_{u,d}$ are not considered as the true elementary flavor-breaking
structures, there is no reason for the superpotential to be holomorphic in
them. In particular, in view of the expansions in Eq.~(\ref{SoftQ}), one may
consider the extended parametrization%
\begin{equation}
\boldsymbol{\lambda}  ^{\prime\prime IJK}=\lambda\times\varepsilon
^{LMN}(\mathbf{X}_{U}\cdot\gamma\mathbf{Y}_{u}\cdot\mathbf{X}_{Q}%
)^{IL}(\mathbf{X}_{D}\cdot\gamma\mathbf{Y}_{D}\cdot\mathbf{X}_{Q}%
)^{JM}(\mathbf{X}_{D}\cdot\gamma\mathbf{Y}_{D}\cdot\mathbf{X}_{Q})^{KN}\;,
\end{equation}
where $\lambda$ and $\gamma$ are numerical factors. In the spirit of the
previous section, one could think such a term would arise if $\Delta
\mathcal{B}$ occurs only in the (holomorphic) couplings between some new
states. It is then communicated to (s)quarks through their tree-level exchanges.

Phenomenologically, this parametrization collapses to the one in
Eq.~(\ref{Holo1}). First, the contraction of the three $\mathbf{X}_{Q}$ can be
simplified using $\varepsilon^{LMN}A^{IL}A^{JM}A^{KN}=\varepsilon^{IJK}\det A$
as
\begin{equation}
\boldsymbol{\lambda}  ^{\prime\prime IJK}=\lambda\times\gamma^{3}%
\det(\mathbf{X}_{Q})\times\varepsilon^{LMN}(\mathbf{X}_{U}\cdot\mathbf{Y}%
_{u})^{IL}(\mathbf{X}_{D}\cdot\mathbf{Y}_{D})^{JM}(\mathbf{X}_{D}%
\cdot\mathbf{Y}_{D})^{KN}\;.
\end{equation}
Then, if this structure arise at the high-scale, it will run down towards
Eq.~(\ref{Holo1}) thanks to its attractor property~\cite{BernonS14}. For
example, if one starts with the factor
\begin{equation}
\mathbf{X}_{U}=\mathbf{1}+c\mathbf{Y}_{u}^{\dagger}\mathbf{Y}_{u}%
\;,\;\;c[M_{GUT}]=-\frac{1}{\langle\mathbf{Y}_{u}^{\dagger}\mathbf{Y}%
_{u}\rangle}\sim\mathcal{O}(1)\;,
\end{equation}
one ends up with $c[1\,TeV]\sim\mathcal{O}(10^{-2})$. The geometric
suppression of $(\mathbf{X}_{U,D})_{33}$ is wiped out by the RG evolution, and
one effectively remains with only the holomorphic term Eq.~(\ref{Holo1}).

The only issue is thus the overall size of the $\boldsymbol{\lambda}
^{\prime\prime}$ coupling, which has to be sufficient to prevent LSP squarks
or gluinos to be long-lived. The whole RG evolution~\cite{BernonS14} from the
GUT to the TeV scale amount to reducing $\lambda$ by about a factor 5. So, the
factor $\gamma^{3}\det(\mathbf{X}_{Q})$ should be of $\mathcal{O}(1)$, which
requires $\gamma$ to be $\mathcal{O}(10)$ to compensate for the strong
suppression of $\det(\mathbf{X}_{Q})$ when $\eta_{q}$ becomes large, see
Eq.~(\ref{SoftXDUD}). This value of $\gamma$ is, coincidentally, very close to
that found in Eq.~(\ref{SleptRG}) when imposing $\mathbf{Y}_{e}=\gamma
\mathbf{Y}_{d}\cdot\mathbf{X}_{Q}$.

In conclusion, the R-parity violating sector is not affected significantly by
geometric MFV expansions, and thus retains all its capabilities at hiding
low-scale supersymmetry.

\section{Scenario 3: Minimal SU(5) with true quark-lepton unification\label{SecGUT}}

When discussing unification of quarks and leptons, GUTs immediately jump to
mind, so it is now time to analyze how the strategy developed in the previous
sections translate in such settings. In Sec.~\ref{Sec51}, we first recall how
the flavor sector of the minimal SU(5) model is constructed (see e.g.
Ref.~\cite{Langacker:1980js} for a review), along with the standard strategies
aimed at correcting its prediction $\mathbf{Y}_{e}^{T}=\mathbf{Y}_{d}$. Then,
in Sec.~\ref{Sec52}, we show how geometric MFV expansions can help
resolve the quark-lepton unification puzzle of $SU(5)$ in a minimal and
natural way.

\subsection{Flavor disunification in minimal unification models\label{Sec51}}

In the minimal SU(5) unification model, the quarks and leptons are embedded
into the $\mathbf{5}$ and $\mathbf{10}$ representations, denoted
$\psi_{\mathbf{\bar{5}}}=\bar{D}\oplus L$ and $\chi_{\mathbf{10}}%
=Q\oplus\bar{U}\oplus\bar{E}$. Their SU(5)-symmetric Yukawa couplings are%
\begin{equation}
\mathcal{L}_{\text{Yukawa}}=-\frac{1}{4}\bar{\chi}_{\mathbf{10}}^{\mathrm{C}%
}\mathbf{Y}_{10}\chi_{\mathbf{10}}h_{\mathbf{5}}^{u}+\sqrt{2}\bar{\psi
}_{\mathbf{\bar{5}}}^{\mathrm{C}}\mathbf{Y}_{5}\chi_{\mathbf{10}%
}h_{\mathbf{\bar{5}}}^{d}+h.c.\;, \label{L5}%
\end{equation}
where $\mathrm{C}$ stands for charge conjugation. After the spontaneous
breaking of $SU(5)$ down to $SU(3)_{C}\otimes SU(2)_{L}\otimes U(1)_{Y}$
through the adjoint Higgs field $H_{24}$, these couplings split into the usual
quark and lepton Yukawa couplings of the THDM of type II , with the matching
conditions at the GUT scale
\begin{equation}
\left\{
\begin{array}
[c]{l}%
\mathbf{Y}_{u}=\mathbf{Y}_{10}\;,\;\;[\mathbf{Y}_{10}=\mathbf{Y}_{10}%
^{T}]\;\;\smallskip\\
\mathbf{Y}_{d}=\mathbf{Y}_{5}\;,\smallskip\\
\mathbf{Y}_{e}=\mathbf{Y}_{5}^{T}\;.
\end{array}
\right.
\end{equation}
Charged lepton and down-type quark masses are thus equal at the unification
scale, $m_{e}=m_{d}$, $m_{\mu}=m_{s}$, and $m_{\tau}=m_{b}$. At the EW scale,
the neutral components of the $h_{\mathbf{5}}^{u}$ and $h_{\mathbf{\bar{5}}%
}^{d}$ fields break $SU(3)_{C}\otimes SU(2)_{L}\otimes U(1)_{Y}$ down to
$SU(3)_{C}\otimes U(1)_{em}$, and accounting for the rather fast QCD evolution
of the quark masses, one gets%
\begin{equation}
m_{b}\approx3m_{\tau}\;,\;\;\frac{m_{d}}{m_{s}}\approx\frac{m_{e}}{m_{\mu}}\;.
\end{equation}
The first relation is rather well satisfied but the second is badly violated,
$1/20\not \approx 1/200$.

There are two well-known ways to improve the mass ratios. The first is to
introduce a set of scalar fields transforming in the $\mathbf{45}$
representation~\cite{GJ45}. The additional Yukawa couplings have the explicit
form
\begin{equation}
\mathcal{L}_{\text{Yukawa}}^{\mathbf{45}}=\sqrt{3/8}\bar{\chi}_{\mathbf{10}%
}^{\mathrm{C}}\mathbf{Y}_{10}^{\prime}\chi_{\mathbf{10}}h_{\mathbf{45}}%
^{u}-\sqrt{12}\bar{\psi}_{\mathbf{\bar{5}}}^{\mathrm{C}}\mathbf{Y}_{5}%
^{\prime}\chi_{\mathbf{10}}h_{\overline{\mathbf{45}}}^{d}+h.c.\;. \label{L45}%
\end{equation}
After the $SU(5)$ and EW symmetry breaking, now induced by the four scalar
fields $h_{\mathbf{5},\mathbf{45}}^{u}$ and $h_{\mathbf{\bar{5}}%
,\overline{\mathbf{45}}}^{d}$, the matching with the low-scale Yukawa
couplings become%
\begin{equation}
\left\{
\begin{array}
[c]{l}%
\mathbf{Y}_{u}=\mathbf{Y}_{10}\sin\alpha_{u}+\mathbf{Y}_{10}^{\prime}%
\cos\alpha_{u}\;,\;\;[\mathbf{Y}_{10}=\mathbf{Y}_{10}^{T},\;\mathbf{Y}%
_{10}^{\prime}=-\mathbf{Y}_{10}^{\prime T}]\;\smallskip\\
\mathbf{Y}_{d}=\mathbf{Y}_{5}\sin\alpha_{d}+\mathbf{Y}_{5}^{\prime}\cos
\alpha_{d}\;,\;\smallskip\\
\mathbf{Y}_{e}=\mathbf{Y}_{5}^{T}\sin\alpha_{d}-3\mathbf{Y}_{5}^{\prime T}%
\cos\alpha_{d}\;,
\end{array}
\right.  \label{Yuk45}%
\end{equation}
where $\tan\alpha_{i}=v_{5}^{i}/v_{45}^{i}$ and $v_{\mathbf{r}}^{i}$ the VEV
of the neutral $h_{\mathbf{r}}^{i}$ components. The second path to cure the
mass ratios is to keep the scalar content minimal but allow for
higher-dimensional Yukawa couplings. The possible dimension-five couplings
are:
\begin{align}
\mathcal{L}_{\text{Yukawa}}^{\dim-5}  &  =\frac{\sqrt{2}}{\Lambda}\bar{\psi
}_{\mathbf{\bar{5}}}^{\mathrm{C}}\mathbf{Y}_{5}^{\prime}(\mathbf{H}%
_{\mathbf{24}}\chi_{\mathbf{10}})h_{\mathbf{\bar{5}}}^{d}-\frac{2}{\Lambda
}\bar{\chi}_{\mathbf{10}}^{\mathrm{C}}\mathbf{Y}_{10}^{\prime}(\chi
_{\mathbf{10}}\mathbf{H}_{\mathbf{24}})h_{\mathbf{5}}^{u}\nonumber\\
&  \;\;\;\;-\frac{1}{\Lambda}\bar{\psi}_{\mathbf{\bar{5}}}^{\mathrm{C}%
}\mathbf{Y}_{5}^{\prime\prime}\chi_{\mathbf{10}}(h_{\mathbf{\bar{5}}}^{d}\mathbf{H}_{\mathbf{24}})+\frac{1}{\Lambda}\bar{\chi}_{\mathbf{10}}^{\mathrm{C}}\mathbf{Y}_{10}^{\prime\prime}\chi_{\mathbf{10}}(h_{\mathbf{5}}^{u}\mathbf{H}_{\mathbf{24}})\;. \label{Ldim5}%
\end{align}
Writing $SU(5)$ indices explicitly, the combinations $(h_{\mathbf{\bar{5}}}^{d} \mathbf{H}_{\mathbf{24}})=(h_{\mathbf{\bar{5}}}^{d})_{B}(\mathbf{H}_{\mathbf{24}})_{A}^{B}$ and $(h_{\mathbf{5}}^{u}\mathbf{H}_{\mathbf{24}})=(h_{\mathbf{5}}^{u})^{B}(\mathbf{H}_{\mathbf{24}})_{B}^{A}$ transform as  $\mathbf{\bar{5}}$ and $\mathbf{5}$, respectively, so the $\mathbf{Y}_{5}^{\prime\prime}$ and $\mathbf{Y}_{10}^{\prime\prime}$ couplings can be
absorbed into $\mathbf{Y}_{5}$ and $\mathbf{Y}_{10}$ of Eq.~(\ref{L5}). For
the other two couplings, $(h_{\mathbf{\bar{5}}}^{d})_{A}(\mathbf{H}_{\mathbf{24}})_{C}^{B}$ and $(h_{\mathbf{5}}^{u})^{A}(\mathbf{H}_{\mathbf{24}})_{C}^{B}$ contain in addition a piece transforming like $\overline{\mathbf{45}}$ and $\mathbf{45}$, respectively, which thus acts like the extra scalar fields of Eq.~(\ref{L45}). Explicitly, the low-scale Yukawa couplings become%
\begin{equation}
\left\{
\begin{array}
[c]{l}%
\mathbf{Y}_{u}=\mathbf{Y}_{10}+\lambda(4\mathbf{Y}_{10}^{\prime T}%
-\mathbf{Y}_{10}^{\prime})\;,\;\;[\mathbf{Y}_{10}=\mathbf{Y}_{10}%
^{T}]\smallskip\\
\mathbf{Y}_{d}=\mathbf{Y}_{5}+\lambda\mathbf{Y}_{5}^{\prime}\;,\;\smallskip\\
\mathbf{Y}_{e}=\mathbf{Y}_{5}^{T}-\frac{3}{2}\lambda\mathbf{Y}_{5}^{\prime
T}\;,
\end{array}
\right.  \label{YukDim5}%
\end{equation}
where $\lambda=v_{24}/\Lambda$.

Even if correct mass ratios are trivially obtained, these strategies are not
satisfactory from a flavor point of view. First, they both fail to truly unify
quarks and leptons since additional flavor structures have to be introduced.
Second, in a supersymmetric context, FCNC are not necessarily under control.
To understand this last point, remark first that the flavor group is
$G_{F}=U(3)_{\chi_{\mathbf{10}}}\otimes U(3)_{\psi_{\mathbf{\bar{5}}}}$ at the
GUT level. If only $\mathbf{Y}_{10}$ and $\mathbf{Y}_{5}$ are spurions, this
is sufficient to bring them to their background values%
\begin{equation}
\mathbf{Y}_{10}\rightarrow\mathbf{D}_{10}\;,\;\mathbf{Y}_{5}\rightarrow
\mathbf{D}_{5}U^{\dagger}\;, \label{SU5Back}%
\end{equation}
where the real diagonal matrices $\mathbf{D}_{10}$ and $\mathbf{D}_{5}$ are
defined from the decompositions%
\begin{equation}
V_{5}^{\ast}\mathbf{Y}_{5}V_{10}^{\prime\dagger}=\mathbf{D}_{5}\;\;,\;\;V_{10}%
^{\ast}\mathbf{Y}_{10}V_{10}^{\dagger}=\mathbf{D}_{10}\;, \label{SU5SVD}%
\end{equation}
and with $U\equiv V_{10}V_{10}^{\prime\dagger}$. In the absence of any other
spurion, $\mathbf{D}_{10}=\operatorname{diag}(y_{u},y_{c},y_{t})$,
$\mathbf{D}_{5}=\operatorname{diag}(y_{e}=y_{d},y_{\mu}=y_{s},y_{\tau}=y_{b}%
)$, and $U$ is equal to the CKM matrix up to two Majorana phases.

Adding spurions like $\mathbf{Y}_{5}^{\prime}$ or $\mathbf{Y}_{10}^{\prime}$
to this list, the flavor group is no longer large enough to bring all of them
to their background values, and unknown mixing matrices remain~\cite{GUTMFV}.
Specifically, the unitary rotations of the fermion fields is defined from the
SVD of the $\mathbf{Y}_{u}$, $\mathbf{Y}_{d}$, and $\mathbf{Y}_{e}$ couplings,
with those now given by the combinations in Eq.~(\ref{Yuk45})
or~(\ref{YukDim5}). This permits to reach the basis in Eq.~(\ref{PhysV1}). The
same unitary rotations have to be performed on the sfermion partners. But,
consider the sfermion soft-terms%
\begin{equation}
\mathcal{L}_{soft}\ni-(\mathbf{m}_{\mathbf{10}}^{2})\langle\tilde{\chi
}_{\mathbf{10}}^{\dagger}\tilde{\chi}_{\mathbf{10}}\rangle-(\mathbf{m}%
_{\mathbf{5}}^{2})\tilde{\psi}_{\mathbf{\bar{5}}}^{\dagger}\tilde{\psi
}_{\mathbf{\bar{5}}}\;,
\end{equation}
which take the generic form%
\begin{align}
\mathbf{m}_{\mathbf{10}}^{2}  &  =c_{0}\mathbf{1}+c_{1}\mathbf{Y}%
_{10}^{\dagger}\mathbf{Y}_{10}+c_{2}\mathbf{Y}_{5}^{\dagger}\mathbf{Y}%
_{5}+c_{3}\mathbf{Y}_{10}^{\dagger}\mathbf{Y}_{10}^{\prime}+c_{4}%
\mathbf{Y}_{5}^{\dagger}\mathbf{Y}_{5}^{\prime}+\mathbf{...\;,}\\
\mathbf{m}_{\mathbf{5}}^{2}  &  =c_{0}\mathbf{1}+c_{1}\mathbf{Y}_{5}^{\ast
}\mathbf{Y}_{5}^{T}+c_{2}\mathbf{Y}_{5}^{\ast}\mathbf{Y}_{5}^{\prime T}%
+c_{3}\mathbf{Y}_{5}^{\ast}\mathbf{Y}_{10}^{T}\mathbf{Y}_{10}^{\ast}%
\mathbf{Y}_{5}^{T}+c_{4}\mathbf{Y}_{5}^{\ast}\mathbf{Y}_{10}^{T}%
\mathbf{Y}_{10}^{\prime\ast}\mathbf{Y}_{5}^{T}+\mathbf{...\;,}%
\end{align}
for some $c_{i}$ coefficients. Rotating the sfermions does not permit in general
to reach a form where $\mathbf{m}_{\mathbf{10}}^{2}$ and $\mathbf{m}%
_{\mathbf{5}}^{2}$ are entirely given out of the fermion masses and CKM
matrix, because the action of the SVD unitary matrices is only known for the
specific combinations in Eq.~(\ref{Yuk45}) or~(\ref{YukDim5}), and not
individually on $\mathbf{Y}_{5}$, $\mathbf{Y}_{5}^{\prime}$, $\mathbf{Y}_{10}%
$, and $\mathbf{Y}_{10}^{\prime}$. Unknown unitary matrices remain, the
sfermion soft-terms are a priori far from their MFV form, and when run down,
generate potentially devastating contributions to FCNC.

\subsection{Towards dynamical flavor unification\label{Sec52}}

We know from the previous sections that $\mathbf{Y}_{e}$ can be expressed in
terms of $\mathbf{Y}_{u}$ and $\mathbf{Y}_{d}$, so the same must be true in
the context of $SU(5)$. It must be possible to express the flavor structures
$\mathbf{Y}_{5}^{\prime}$ and $\mathbf{Y}_{10}^{\prime}$ coming from either
Eq.~(\ref{L45}) or~(\ref{Ldim5}) as expansions in $\mathbf{Y}_{5}$ and
$\mathbf{Y}_{10}$, and still get correct mass ratios at the GUT scale. The
whole flavor structure of the model, even in a supersymmetric context, would
then be fixed entirely in terms of only two spurions, themselves fixed from
the known fermion masses and CKM mixing.

Of course, since a finite polynomial relationship between $\mathbf{Y}_{e}$ and
$\mathbf{Y}_{u,d}$ is necessarily fine-tuned, so are a priori those relating
$\mathbf{Y}_{5}^{\prime}$ and $\mathbf{Y}_{10}^{\prime}$ to $\mathbf{Y}_{5}$
and $\mathbf{Y}_{10}$. Infinite series are again compulsory. To illustrate
this in a realistic setting, let us construct a model inspired from that in
Section~\ref{SecVL}. We introduce flavor triplets of vector-like fermions, here
transforming as $X_{\mathbf{10}}^{L,R}\sim\mathbf{10}$, $X_{\mathbf{\bar{5}}%
}^{L,R}\sim\mathbf{\bar{5}}$. To the $SU(5)$ Yukawa couplings%
\begin{equation}
\mathcal{L}_{\text{Yukawa}}^{\text{Chiral}}=-\frac{1}{4}\bar{\chi
}_{\mathbf{10}}^{\mathrm{C}}\mathbf{Y}_{10}\chi_{\mathbf{10}}h_{\mathbf{5}%
}^{u}+\sqrt{2}\bar{\psi}_{\mathbf{\bar{5}}}^{\mathrm{C}}\mathbf{Y}_{5}%
\chi_{\mathbf{10}}h_{\mathbf{\bar{5}}}^{d}+h.c.\;,
\end{equation}
we add%
\begin{align}
\mathcal{L}_{\text{Yukawa}}^{\text{Vector}}  &  =\bar{X}_{\mathbf{\bar{5}}%
}^{L}(\mathbf{M}_{5}+\mathbf{N}_{5}H_{\mathbf{24}})X_{\mathbf{\bar{5}}}%
^{R}+\bar{X}_{\mathbf{10}}^{L}(\mathbf{M}_{10}+\mathbf{N}_{10}H_{\mathbf{24}%
})X_{\mathbf{10}}^{R}\nonumber\\
&  \;\;\;\;-\frac{1}{4}\bar{X}_{\mathbf{10}}^{L,\,\mathrm{C}}\mathbf{Y}%
_{10}^{L}X_{\mathbf{10}}^{L}h_{\mathbf{5}}^{u}-\frac{1}{4}\bar{X}%
_{\mathbf{10}}^{R,\,\mathrm{C}}\mathbf{Y}_{10}^{R}X_{\mathbf{10}}%
^{R}h_{\mathbf{5}}^{u}\nonumber\\
&  \;\;\;\;+\sqrt{2}\bar{X}_{\mathbf{\bar{5}}}^{L,\mathrm{C}}\mathbf{Y}%
_{5}^{L}X_{\mathbf{10}}^{L}h_{\mathbf{\bar{5}}}^{d}+\sqrt{2}\bar
{X}_{\mathbf{\bar{5}}}^{R,\mathrm{C}}\mathbf{Y}_{5}^{R}X_{\mathbf{10}}%
^{R}h_{\mathbf{\bar{5}}}^{d}+h.c.\;,
\end{align}
where flavor indices are suppressed, as well as mixed Yukawa interactions
\begin{align}
\mathcal{L}_{\text{Yukawa}}^{\text{Mixed}}  &  =\frac{1}{2}\bar{\chi
}_{\mathbf{10}}^{\mathrm{C}}\mathbf{\bar{Y}}_{10}X_{\mathbf{10}}%
^{L}h_{\mathbf{5}}^{u}+\sqrt{2}\bar{\psi}_{\mathbf{\bar{5}}}^{\mathrm{C}%
}\mathbf{\bar{Y}}_{5}X_{\mathbf{10}}^{L}h_{\mathbf{\bar{5}}}^{d}+\sqrt{2}%
\bar{X}_{\mathbf{\bar{5}}}^{L,\mathrm{C}}\mathbf{\bar{Y}}_{5}\chi
_{\mathbf{10}}h_{\mathbf{\bar{5}}}^{d}\nonumber\\
&  \;\;\;\;+\bar{X}_{\mathbf{10}}^{R}\mathbf{\bar{N}}_{10}H_{\mathbf{24}}%
\chi_{\mathbf{10}}+\bar{X}_{\mathbf{\bar{5}}}^{R}\mathbf{\bar{N}}%
_{5}H_{\mathbf{24}}\psi_{\mathbf{\bar{5}}}+h.c.\;,
\end{align}
where we have identified the $\bar{\psi}_{\mathbf{\bar{5}}}^{\mathrm{C}%
}\mathbf{\bar{Y}}_{5}X_{\mathbf{10}}^{L}h_{\mathbf{\bar{5}}}^{d}$ and
$\sqrt{2}\bar{X}_{\mathbf{\bar{5}}}^{L,\mathrm{C}}\mathbf{\bar{Y}}_{5}%
\chi_{\mathbf{10}}h_{\mathbf{\bar{5}}}^{d}$ couplings for simplicity. Compared
to the vector-like fermion model in Section~\ref{SecVL}, there is no need to introduce a
singlet Higgs field. The adjoint Higgs boson $H_{24}$ with its very large VEV
$v_{24}$ can perfectly take its place.

When the heavy fermions are integrated out, an infinite tower of effective
Yukawa couplings for $\chi_{\mathbf{10}}$ and $\psi_{\mathbf{\bar{5}}}$ are
generated, starting with the five-dimensional operators of Eq.~(\ref{Ldim5}).
Upon enforcing MFV under the flavor group%
\begin{equation}
G_{F}=U(3)_{\chi_{\mathbf{10}}=X_{\mathbf{10}}^{L,R}}\otimes U(3)_{\psi
_{\mathbf{\bar{5}}}=X_{\mathbf{5}}^{L,R}}\;,
\end{equation}
on the new couplings in $\mathcal{L}_{\text{Yukawa}}^{\text{Vector}}$ and
$\mathcal{L}_{\text{Yukawa}}^{\text{Mixed}}$, for instance as
\begin{subequations}
\label{MFVSU5}%
\begin{align}
&  \left\{
\begin{array}
[c]{l}%
\mathbf{M}_{5}=M_{5}\mathbf{1}\;,\smallskip\\
\mathbf{Y}_{5}=\mathbf{Y}_{5}^{L}=\mathbf{Y}_{5}^{R}=\mathbf{\bar{Y}}%
_{5}\;,\smallskip\\
\mathbf{N}_{5}=\mathbf{\bar{N}}_{5}=\alpha_{1}\mathbf{Y}_{5}^{\ast}%
\mathbf{Y}_{5}^{T}+\alpha_{2}\mathbf{Y}_{5}^{\ast}\mathbf{Y}_{10}%
^{T}\mathbf{Y}_{10}^{\ast}\mathbf{Y}_{5}^{T}+\alpha_{3}\mathbf{Y}_{5}^{\ast
}\mathbf{Y}_{5}^{T}\mathbf{Y}_{5}^{\ast}\mathbf{Y}_{5}^{T}+...\;,
\end{array}
\right.  \;\\
&  \left\{
\begin{array}
[c]{l}%
\mathbf{M}_{10}=M_{10}\mathbf{1}\;,\smallskip\\
\mathbf{Y}_{10}=\mathbf{Y}_{10}^{L}=\mathbf{Y}_{10}^{R}=\mathbf{\bar{Y}}%
_{10}\;,\smallskip\\
\mathbf{N}_{10}=\mathbf{\bar{N}}_{10}=\beta_{1}\mathbf{Y}_{10}^{\dagger
}\mathbf{Y}_{10}+\beta_{2}\mathbf{Y}_{5}^{\dagger}\mathbf{Y}_{5}+\beta
_{3}\mathbf{Y}_{10}^{\dagger}\mathbf{Y}_{10}\mathbf{Y}_{10}^{\dagger
}\mathbf{Y}_{10}+...\;,
\end{array}
\right.
\end{align}
these higher-dimensional effective interactions will automatically be
expressed in terms of $\mathbf{Y}_{5}$ and $\mathbf{Y}_{10}$. Note that these
MFV conditions can be understood in the same way as in Eq.~(\ref{U91}).
Arranging the fields transforming identically under the gauge group into
$\Psi_{\mathbf{10}}=(X_{\mathbf{10}}^{L},\chi_{\mathbf{10}},X_{\mathbf{10}%
}^{R})$ and $\Psi_{\mathbf{5}}=(X_{\mathbf{\bar{5}}}^{L},\psi_{\mathbf{\bar
{5}}},X_{\mathbf{\bar{5}}}^{R})$, the structure of the model matches
Eq.~(\ref{U92}), up to obvious substitutions.

The set of coupled equations of motion can be solved iteratively, though it is
quite cumbersome and moved to the Appendix. Only the leading order in
$h_{\mathbf{5}}^{u}$ and $h_{\mathbf{\bar{5}}}^{d}$ needs to be kept since
$v_{5}^{u,d}/v_{24}\ll1$, in which case this iterative procedure quickly
terminates. After the $SU(5)$ breaking, the effective Yukawa couplings are
found to be
\end{subequations}
\begin{equation}
\left\{
\begin{array}
[c]{l}%
\mathbf{Y}_{u}=\mathbf{Y}_{10}-\mathbf{\bar{Y}}_{10}\cdot\mathbf{F}%
_{\mathbf{10}}^{-1/4,T}-\mathbf{F}_{\mathbf{10}}^{1}\cdot\mathbf{\bar{Y}}%
_{10}+\mathbf{F}_{\mathbf{10}}^{1}\cdot\mathbf{Y}_{10}^{L}\cdot\mathbf{F}%
_{\mathbf{10}}^{-1/4,T}\;,\smallskip\\
\mathbf{Y}_{d}=\mathbf{Y}_{5}-\mathbf{F}_{\mathbf{5}}^{1}\cdot\mathbf{\bar{Y}%
}_{5}-\mathbf{\bar{Y}}_{5}\cdot\mathbf{F}_{\mathbf{10}}^{-1/4,T}%
+\mathbf{F}_{\mathbf{5}}^{1}\cdot\mathbf{Y}_{5}^{L}\cdot\mathbf{F}%
_{\mathbf{10}}^{-1/4,T}\;,\smallskip\\
\mathbf{Y}_{e}^{T}=\mathbf{Y}_{5}-\mathbf{F}_{\mathbf{5}}^{-3/2}%
\cdot\mathbf{\bar{Y}}_{5}-\mathbf{\bar{Y}}_{5}\cdot\mathbf{F}_{\mathbf{10}%
}^{-3/2,T}+\mathbf{F}_{\mathbf{5}}^{-3/2}\cdot\mathbf{Y}_{5}^{L}%
\cdot\mathbf{F}_{\mathbf{10}}^{-3/2,T}\;,
\end{array}
\right.  \label{EffectiveSU5Yuk}%
\end{equation}
with $\mathbf{F}_{\mathbf{R}}^{\alpha}=\alpha v_{24}\mathbf{N}_{\mathbf{R}%
}^{M}\cdot(\mathbf{M}_{\mathbf{R}}+\alpha v_{24}\mathbf{N}_{\mathbf{R}})^{-1}%
$. Upon enforcing the MFV conditions in Eq.~(\ref{MFVSU5}), this becomes%
\begin{equation}
\left\{
\begin{array}
[c]{l}%
\mathbf{Y}_{u}\overset{MFV}{=}(1-\mathbf{F}_{\mathbf{10}}^{1})\cdot
\mathbf{Y}_{10}\cdot(1-\mathbf{F}_{\mathbf{10}}^{-1/4,T})\;,\smallskip\\
\mathbf{Y}_{d}\overset{MFV}{=}(1-\mathbf{F}_{\mathbf{5}}^{1})\cdot
\mathbf{Y}_{5}\cdot(1-\mathbf{F}_{\mathbf{10}}^{-1/4,T})\;,\smallskip\\
\mathbf{Y}_{e}^{T}\overset{MFV}{=}(1-\mathbf{F}_{\mathbf{5}}^{-3/2}%
)\cdot\mathbf{Y}_{5}\cdot(1-\mathbf{F}_{\mathbf{10}}^{-3/2,T})\;,
\end{array}
\right.  \label{EffSU5MFV}%
\end{equation}
while%
\begin{equation}
1-\mathbf{F}_{\mathbf{R}}^{\alpha}\overset{MFV}{=}\frac{1}{1+\alpha
\frac{v_{24}}{M_{R}}\mathbf{N}_{\mathbf{R}}}\;.
\end{equation}
We thus recover geometric series, but now for all three Yukawa couplings simultaneously.

Despite their rather simple appearance, these equations are difficult to
solve. First, one should realize that even if $\mathbf{Y}_{10}$ and
$\mathbf{Y}_{5}$ are the only spurions, their background values are unknown.
Using the $G_{F}^{\prime}$ symmetry, these two spurions can be rotated to
$\mathbf{Y}_{10}\rightarrow\mathbf{D}_{10}$ and $\mathbf{Y}_{5}\rightarrow
\mathbf{D}_{5}U^{\dagger}$, but $\mathbf{D}_{10}$ and $\mathbf{D}_{5}$ are not
simply given in terms of the observed quark masses, and $U$ is not equal the
CKM matrix. This leaves the six diagonal entries and the six parameters
entering $U$ as free parameters. To this, we should add the free parameters
entering $\mathbf{F}_{\mathbf{R}}^{\alpha}$, which are the two vector fermion
masses, and the expansion parameters in $\mathbf{N}_{\mathbf{R}}$. All these
free parameters must be fixed so that the SVD of $\mathbf{Y}_{u}$,
$\mathbf{Y}_{d}$, and $\mathbf{Y}_{e}$ reproduce Eq.~(\ref{PhysV1}), that is,
the nine singular values have to match the quark and lepton masses, and the
mismatch between the left SVD unitary matrices for $\mathbf{Y}_{u}$ and
$\mathbf{Y}_{d}$ has to reproduce precisely the CKM matrix. To add to the
difficulty, these equations are highly non-linear, so there will be many
solutions, but we are after those making most sense physically. That is, we
want the coefficients in $\mathbf{N}_{\mathbf{R}}$ to be $\mathcal{O}(1)$, and
the solution to be rather stable against small variations in these
coefficients, $\mathbf{Y}_{10}$ and $\mathbf{Y}_{5}$ entries, or $v_{24}%
/M_{R}$.

Solving these equations in the general case represents a formidable task which
we leave for future works. Rather, let us go back to the issue of the relative
normalization between $\mathbf{Y}_{d}$ and $\mathbf{Y}_{e}$, which had to be
tuned by a free parameter in Section~\ref{SecVL}, see Eq.~(\ref{CondX}) or~(\ref{CondXZ}%
), but is now imposed by the $SU(5)$ symmetry. To this end, we solve the
system of equations under the approximation that $\mathbf{Y}_{10}$ and
$\mathbf{Y}_{5}$ are simultaneously diagonal, and consider separately the
exchange of either $X_{\mathbf{5}}^{L,R}$ or $X_{\mathbf{10}}^{L,R}$. For the
former case, we find for the MSSM at $\tan\beta=50$,%
\begin{gather}
\mathbf{Y}_{10}=\operatorname{diag}(y_{u},y_{c},y_{t})\;,\;\mathbf{Y}%
_{5}=\left(
\begin{array}
[c]{ccc}%
0.0073 & 0 & 0\\
0 & 0.00016 & 0\\
0 & 0 & 0.39
\end{array}
\right)  \;,\;\nonumber\\
\frac{v_{24}}{M_{5}}=1.5\times10^{7}\;,\;\;\alpha_{1}\equiv-1\;,\;\alpha
_{2}=4.8\;,\;\alpha_{3}=6.7\;,
\end{gather}
producing $\mathbf{Y}_{d}=\operatorname{diag}(y_{s},y_{d},y_{b})$ and
$\mathbf{Y}_{e}=\operatorname{diag}(y_{\mu},y_{e},y_{\tau})$. If we instead
keep only the six $X_{\mathbf{10}}^{L,R}$ fields,%
\begin{gather}
\mathbf{Y}_{10}=\left(
\begin{array}
[c]{ccc}%
-0.000003 & 0 & 0\\
0 & 0.0015 & 0\\
0 & 0 & -1.58
\end{array}
\right)  \;,\;\mathbf{Y}_{5}=\left(
\begin{array}
[c]{ccc}%
0.00039 & 0 & 0\\
0 & 0.0042 & 0\\
0 & 0 & -0.17
\end{array}
\right)  \;,\;\nonumber\\
\frac{v_{24}}{M_{10}}=1.2\times10^{7}\;,\;\;\beta_{1}\equiv-1\;,\;\beta
_{2}=-7.6\;,\;\beta_{3}=-3.0\;,
\end{gather}
then producing $\mathbf{Y}_{u}=\operatorname{diag}(y_{u},y_{c},y_{t})$,
$\mathbf{Y}_{d}=\operatorname{diag}(y_{d},y_{s},y_{b})$, and $\mathbf{Y}%
_{e}=\operatorname{diag}(y_{e},y_{\tau},y_{\mu})$. These solutions share a
number of characteristics:

\begin{itemize}
\item Both generate acceptable coefficients, with the vector-fermion scale
coincidentally close to the usual neutrino seesaw scale at around $10^{9}$ GeV.

\item The hierarchies of $\mathbf{Y}_{5}$ and $\mathbf{Y}_{10}$ end up
aligned, though that of $\mathbf{Y}_{u}$, $\mathbf{Y}_{d}$, and $\mathbf{Y}%
_{e}$ depends on the solution. Mathematically, the system of equations in the
diagonal approximation can be solved whatever the chosen hierarchy for
$\mathbf{Y}_{u}$, $\mathbf{Y}_{d}$, and $\mathbf{Y}_{e}$, and we here present
only two examples. In this respect, the first solution twists the down quarks
and not the up quarks, so it may seem in obvious contradiction with the known
CKM matrix. This is an artifact of the diagonal approximation. Once
$U\neq\mathbf{1}$, only the right-handed down quarks and left-handed leptons
are mixed since the geometric series induced by $X_{\mathbf{\bar{5}}}^{L,R}$
acts on $\psi_{\mathbf{\bar{5}}}=\bar{D}\oplus L$. The hierarchy of the
left-handed down quarks is maintained aligned with that of the up quarks.

\item These solutions exhibit again some serious fine-tuning. The $\mathbf{N}_{5}$ or $\mathbf{N}_{10}$ terms have rather suppressed 3-3 entries, and are quite analogous to that in Eq.~(\ref{FTcancel0}) and~(\ref{FTcancel}).
\end{itemize}

This last point is particularly undesirable, but is the price to pay for the diagonal approximation. We know from Eq.~(\ref{DiagSVD}) that such diagonal settings can lead to such situations. To give another example, consider solving $\mathbf{Y}_{e}=\gamma\mathbf{Y}_{d}\cdot\mathbf{X}_{Q}$ with $\mathbf{X}_{Q}=(\mathbf{1}+\eta(\mathbf{Y}_{u}^{\dagger}\mathbf{Y}_{u}+\beta\mathbf{Y}_{d}^{\dagger}\mathbf{Y}_{d}))^{-1}$ in the diagonal approximation. Setting $V_{CKM}=\mathbf{1}$, we find the solution $\gamma=7$, $\beta=-1.7$, $\eta=10^{7}$, not so different from Eq.~(\ref{InGUT1}), but here extremely fine-tuned. Changing $\beta$ by as little as $5\%$ reduces the electron Yukawa coupling by more than an order of magnitude. This sensitivity of the expansions to the slight misalignment between $\mathbf{Y}_{u}$ and $\mathbf{Y}_{d}$ can be understood from the structure of $\mathbf{Y}_{u}^{\dagger}\mathbf{Y}_{u}$. Even its diagonal entries are seriously affected, with for example,
\begin{equation}
(\mathbf{Y}_{u}^{\dagger}\mathbf{Y}_{u})^{11}=|V_{us}|^{2}m_{u}^{2}%
+|V_{cd}|^{2}m_{c}^{2}+|V_{td}|^{2}m_{t}^{2}\;,
\end{equation}
being entirely dominated by $|V_{td}|^{2}m_{t}^{2}$. Setting $V_{CKM}=\mathbf{1}$ suppresses $(\mathbf{Y}_{u}^{\dagger}\mathbf{Y}_{u})^{11}$ by no less than seven orders of magnitude, and completely alters the behavior of the solutions. Still, compared to Eq.~(\ref{DiagSVD}), the fact that it is here possible to find acceptable values for the coefficients and vector-fermion scale even in this extreme diagonal case is an excellent indication that Eq.~(\ref{EffSU5MFV}) do admit acceptable solutions in the general case.

In conclusion, let us stress that the simple vector-fermion model presented here is certainly not the final word. It must be seen as a generic strategy to unify quarks and leptons without introducing non-minimal flavor structures. The MFV conditions in Eq.~(\ref{MFVSU5}) could be altered, models where $\mathbf{N}_{5}$ and/or $\mathbf{N}_{10}$ are themselves already geometric series in $\mathbf{Y}_{5}$ and $\mathbf{Y}_{10}$ could be constructed, or the effective Yukawa interactions could be generated through the exchange of states with different quantum numbers, maybe even at the loop level.

\section{Conclusion and perspectives}

In this paper, the unification of the quark and lepton flavor structures was thoroughly revisited. Model-independently, we proved that it is possible to express the lepton Yukawa coupling directly as a polynomial expansion in those of the quarks, but that naturality is not automatic. It requires infinite polynomial expansions, with a geometric-like behavior. We then constructed several models in which such polynomial series are generated dynamically, from the tree-level exchanges of heavy vector-like fermions.

The main results of this analysis are

\begin{itemize}
\item Quite generically, the physical electron states $e_{L}^{phys}$ and/or $e_{R}^{phys}$ end up identified as the $\tau_{L}^{gauge}$ and/or $\tau_{R}^{gauge}$ state, respectively. Indeed, the large top mass generates, through a geometric series, a suppression of the mass of its third generation leptonic partner. The bottom quark mass can play a similar role in a Two Higgs Doublet Model at moderate or large $\tan\beta$. Phenomenologically, such a twist of the left and/or right lepton state is not directly observable at the level of the SM, but could have implications for lepton number violating processes or neutrino mass models.

\item In a supersymmetric context, there are two main consequences. First, the scalar partners of the top quark see their masses suppressed by the geometric resummation, in a way completely similar to the electron. This renders the third generation squarks $\tilde{\tau}_{L,R}$ as well as $\tilde{b}_{L}$ much lighter than the others. Interestingly, such natural SUSY-like mass patterns are the most compatible with the absence of supersymmetric signal at colliders. A second consequence is the presence of lepton flavor violation, even if no new flavor structures were introduced at any stage. Indeed, when $\mathbf{Y}_{e}$ is a function of $\mathbf{Y}_{d,u}$, so are the slepton soft-breaking terms. They thus have non-diagonal entries tuned by the CKM matrix, even in the basis in which $\mathbf{Y}_{e}$ is diagonal. With in addition rather light third generation sleptons, identified as the physical $\Tilde{e}_{L}$ and/or $\tilde{e}_{R}$ states, $\mu\rightarrow e\gamma$ could end up quite close to its current bound.

\item In a GUT context, the same mechanism could in principle be applied. It is thus possible for example within the $SU(5)$ model to have correct unified mass ratios without introducing any additional flavor structure. Phenomenologically, this is most welcome in a supersymmetric setting since it ensures the absence of unknown mixing matrices and their potentially large impact on FCNC. Technically, however, it must be said that inverting the geometric-like expansions of $\mathbf{Y}_{e}$, $\mathbf{Y}_{d}$, and $\mathbf{Y}_{u}$ expressed in terms of $\mathbf{Y}_{5}$, $\mathbf{Y}_{10}$ is particularly tricky, and future work is needed there.
\end{itemize}

These results represent a significant improvement in several respects, but there are still many questions to be resolved. In particular, among the aspects worth studying further, we can mention

\begin{itemize}
\item We have alluded several time at the implications for neutrino models, and these should be studied. The twist identified in the lepton states is certainly a significant new piece of information. At the same time, the situation is quite complicated for neutrinos. If their mass is of the Dirac type, expressing $\mathbf{Y}_{\nu}$ in terms of $\mathbf{Y}_{d,u}$ is mathematically possible but would not be natural since $\mathbf{Y}_{\nu}\ll\mathbf{Y}_{d,u}$. A seesaw mechanism is required to enhance $\mathbf{Y}_{\nu}$ by several orders of magnitude. This necessarily introduce a $\Delta\mathcal{L}=2$ breaking of the flavor symmetry, whose transformation properties are incompatible with those of $\mathbf{Y}_{d,u}$~\cite{MFVBL}. In other words, this breaking term must be part of the spurion content, and the whole analysis becomes far more involved.

\item The dynamical models based on flavor triplets of vector leptons are not final theories and could be improved or modified in many ways. First, the MFV conditions on the vector fermion couplings (see Eq.~(\ref{CondX}),~(\ref{CondXZ}), or~(\ref{MFVSU5})) is not derived from first principles, since the origin of the flavor symmetry and its elementary breaking terms are left to be elucidated. This leaves many alternative boundary conditions to explore. Second, the extended Yukawa sector and the extra scalar state have an impact on how the Peccei-Quinn symmetry is realized and broken, opening the way to fruitful connections with axion models. Third, throughout this work, whenever RG evolution to the GUT scale was performed, the impact of the new states was neglected. This is adequate here since the numerical hierarchies of the flavor couplings would not change much (RGE respect MFV by construction). Still, for the purpose of constructing full models, this approximation should be lifted, especially as vector fermions are known to impact the RGE in a positive way~\cite{VectorUnif}. Fourth, the mass scale of these vector leptons is free, and could actually be quite low, within reach of the LHC. The experimental signatures of such states should be studied further~\cite{VectorLike} because they could offer a direct window into the relationship between quark and lepton mass hierarchies.

\item More generally, vector fermions are not compulsory for our program. Other more complicated settings could be devised, with new states carrying different representations of the gauge group, and contributing at the loop level. Even if the functional dependences between the SM Yukawa couplings would not be as simple as here, the geometric-like behavior needed to naturally relate quark and lepton flavor structures will be reproduced whenever the effective contributions of the new states to the Yukawa couplings decouple when their masses increase.
\end{itemize}

In conclusion, the initial somewhat technical and numerically fine-tuned
relationship between the quark and lepton Yukawa couplings lead us towards a
new generic mechanism, and its accompanying broad range of dynamical
implementations. It opens the way for many applications and extensions, and
truly represents a new paradigm in our quest for quark-lepton unification.

\pagebreak 

\appendix                 

\section{Integrating out SU(5) vector fermions}

First, it is useful to write all the $SU(5)$ indices explicitly. The three pieces of the flavor Lagrangian $\mathcal{L}_{\text{Yukawa}}=\mathcal{L}_{\text{Yukawa}}^{\text{Chiral}%
}+\mathcal{L}_{\text{Yukawa}}^{\text{Vector}}+\mathcal{L}_{\text{Yukawa}%
}^{\text{Mixed}}$ are%
\begin{equation}
\mathcal{L}_{\text{Yukawa}}^{\text{Chiral}}=-\frac{1}{4}\varepsilon
_{ABCDE}(\bar{\chi}_{\mathbf{10}}^{\mathrm{C}})^{AB}\mathbf{Y}_{10}%
(\chi_{\mathbf{10}})^{CD}(h_{\mathbf{5}}^{u})^{E}+\sqrt{2}(\bar{\psi
}_{\mathbf{\bar{5}}}^{\mathrm{C}})_{A}\mathbf{Y}_{5}(\chi_{\mathbf{10}}%
)^{AB}(h_{\mathbf{\bar{5}}}^{d})_{B}+h.c.\;,
\end{equation}
for the chiral fermions,%
\begin{align}
\mathcal{L}_{\text{Yukawa}}^{\text{Vector}}  &  =(\bar{X}_{\mathbf{\bar{5}}%
}^{L})^{A}(\mathbf{M}_{5}\delta_{A}^{B}+\mathbf{N}_{5}(H_{\mathbf{24}}%
)_{A}^{B})(X_{\mathbf{\bar{5}}}^{R})_{B}+(\bar{X}_{\mathbf{10}}^{L}%
)_{AB}(\mathbf{M}_{10}\delta_{C}^{B}+\mathbf{N}_{10}(H_{\mathbf{24}})_{C}%
^{B})(X_{\mathbf{10}}^{R})^{AC}\nonumber\\
&  \;\;\;\;-\frac{1}{4}\varepsilon_{ABCDE}(\bar{X}_{\mathbf{10}}%
^{L,\,\mathrm{C}})^{AB}\mathbf{Y}_{10}^{L}(X_{\mathbf{10}}^{L})^{CD}%
(h_{\mathbf{5}}^{u})^{E}-\frac{1}{4}\varepsilon_{ABCDE}(\bar{X}_{\mathbf{10}%
}^{R,\,\mathrm{C}})^{AB}\mathbf{Y}_{10}^{R}(X_{\mathbf{10}}^{R})^{CD}%
(h_{\mathbf{5}}^{u})^{E}\nonumber\\
&  \;\;\;\;+\sqrt{2}(\bar{X}_{\mathbf{\bar{5}}}^{L,\mathrm{C}})_{A}%
\mathbf{Y}_{5}^{L}(X_{\mathbf{10}}^{L})^{AB}(h_{\mathbf{\bar{5}}}^{d}%
)_{B}+\sqrt{2}(\bar{X}_{\mathbf{\bar{5}}}^{R,\mathrm{C}})_{A}\mathbf{Y}%
_{5}^{R}(X_{\mathbf{10}}^{R})^{AB}(h_{\mathbf{\bar{5}}}^{d})_{B}+h.c.\;,
\end{align}
for the vector fermions, and
\begin{align}
\mathcal{L}_{\text{Yukawa}}^{\text{Mixed}}  &  =-\frac{1}{2}\varepsilon
_{ABCDE}(\bar{\chi}_{\mathbf{10}}^{\mathrm{C}})^{AB}\mathbf{\bar{Y}}%
_{10}(X_{\mathbf{10}}^{L})^{CD}(h_{\mathbf{5}}^{u})^{E}\nonumber\\
&  \;\;\;\;+\sqrt{2}(\bar{\psi}_{\mathbf{\bar{5}}}^{\mathrm{C}})_{A}%
\mathbf{\bar{Y}}_{5}(X_{\mathbf{10}}^{L})^{AB}(h_{\mathbf{\bar{5}}}^{d}%
)_{B}+\sqrt{2}(\bar{X}_{\mathbf{\bar{5}}}^{L,\mathrm{C}})_{A}\mathbf{\bar{Y}%
}_{5}(\chi_{\mathbf{10}})^{AB}(h_{\mathbf{\bar{5}}}^{d})_{B}\\
&  \;\;\;\;+(\bar{X}_{\mathbf{10}}^{R})_{AB}\mathbf{\bar{N}}_{10}%
(\chi_{\mathbf{10}})^{AC}(H_{\mathbf{24}})_{C}^{B}+(\bar{X}_{\mathbf{\bar{5}}%
}^{R})^{B}\mathbf{\bar{N}}_{5}(\psi_{\mathbf{\bar{5}}})_{C}(H_{\mathbf{24}%
})_{B}^{C}\;+h.c.\;,
\end{align}
for the mixed terms, where flavor indices are suppressed.

To extract and solve the equations of motion for the $X_{\mathbf{10}}^{L,R}$
field accounting for their antisymmetry, it is best to first define
\begin{equation}
(\mathbf{\bar{M}}_{10})_{AB}^{CD}=\mathbf{M}_{10}\delta_{AB}^{CD}%
+\mathbf{N}_{10}(H_{\mathbf{24}})_{AB}^{CD}\;,
\end{equation}
with
\begin{subequations}
\begin{equation}
2\delta_{AB}^{CD}\equiv\delta_{A}^{C}\delta_{B}^{D}-\delta_{A}^{D}\delta
_{B}^{C}\;,\;\;4(H_{\mathbf{24}})_{AB}^{CD}\equiv\delta_{A}^{C}(H_{\mathbf{24}%
})_{B}^{D}-\delta_{B}^{C}(H_{\mathbf{24}})_{A}^{D}-\delta_{A}^{D}%
(H_{\mathbf{24}})_{B}^{C}+\delta_{B}^{D}(H_{\mathbf{24}})_{A}^{C}\;,
\end{equation}
so that the couplings take explicitly antisymmetric forms, for example:%
\end{subequations}
\begin{subequations}
\begin{align}
(\bar{X}_{\mathbf{10}}^{R})_{CD}(\mathbf{\bar{M}}_{10})_{AB}^{CD}%
(X_{\mathbf{10}}^{L})^{AB}  &  =(\bar{X}_{\mathbf{10}}^{R})_{AB}%
(\mathbf{M}_{10}\delta_{C}^{B}+\mathbf{N}_{10}(H_{\mathbf{24}})_{C}%
^{B})(X_{\mathbf{10}}^{L})^{AC}\;,\frac{{}}{{}}\\
(\bar{\chi}_{\mathbf{10}})_{CD}\mathbf{\bar{N}}_{10}(X_{\mathbf{10}}^{R}%
)^{AB}(H_{\mathbf{24}})_{AB}^{CD}  &  =(\bar{\chi}_{\mathbf{10}}%
)_{AB}\mathbf{\bar{N}}_{10}(X_{\mathbf{10}}^{R})^{AC}(H_{\mathbf{24}})_{C}%
^{B}\;,\frac{{}}{{}}\\
(\bar{\psi}_{\mathbf{\bar{5}}}^{\mathrm{C}})_{C}\mathbf{\bar{Y}}_{5}%
\delta_{AB}^{CD}(X_{\mathbf{10}}^{L})^{AB}(h_{\mathbf{\bar{5}}}^{d})_{D}  &
=(\bar{\psi}_{\mathbf{\bar{5}}}^{\mathrm{C}})_{A}\mathbf{\bar{Y}}%
_{5}(X_{\mathbf{10}}^{L})^{AB}(h_{\mathbf{\bar{5}}}^{d})_{B}\;.
\end{align}
With this, the equations of motion are%
\end{subequations}
\begin{align}
\frac{\delta\mathcal{L}_{\text{Yukawa}}}{\delta(X_{\mathbf{10}}^{L})^{AB}}  &
=(\bar{X}_{\mathbf{10}}^{R})_{CD}(\mathbf{\bar{M}}_{10})_{AB}^{CD}-\frac{1}%
{2}\varepsilon_{ABCDE}[(\bar{X}_{\mathbf{10}}^{L,\,\mathrm{C}})^{CD}%
\mathbf{Y}_{10}^{L}+(\bar{\chi}_{\mathbf{10}}^{\mathrm{C}})^{CD}%
\mathbf{\bar{Y}}_{10}](h_{\mathbf{5}}^{u})^{E}\nonumber\\
&  \;\;\;+\sqrt{2}[(\bar{X}_{\mathbf{\bar{5}}}^{L,\mathrm{C}})_{C}%
\mathbf{Y}_{5}^{L}+(\bar{\psi}_{\mathbf{\bar{5}}}^{\mathrm{C}}%
)_{C}\mathbf{\bar{Y}}_{5}](h_{\mathbf{\bar{5}}}^{d})_{D}\delta_{AB}%
^{CD}\;,\frac{{}}{{}}\\
\frac{\delta\mathcal{L}_{\text{Yukawa}}}{\delta(X_{\mathbf{10}}^{R})^{AB}}  &
=(\bar{X}_{\mathbf{10}}^{L})_{CD}(\mathbf{\bar{M}}_{10})_{AB}^{CD}+(\bar{\chi
}_{\mathbf{10}})_{CD}\mathbf{\bar{N}}_{10}(H_{\mathbf{24}})_{AB}%
^{CD}\nonumber\\
&  \;\;\;-\frac{1}{2}\varepsilon_{ABCDE}(\bar{X}_{\mathbf{10}}^{R,\,\mathrm{C}%
})^{CD}(h_{\mathbf{5}}^{u})^{E}\mathbf{Y}_{10}^{R}+\sqrt{2}(\bar
{X}_{\mathbf{\bar{5}}}^{R,\mathrm{C}})_{C}(h_{\mathbf{\bar{5}}}^{d})_{D}%
\delta_{AB}^{CD}\mathbf{Y}_{5}^{R}\;.
\end{align}
Those for the $X_{\mathbf{\bar{5}}}^{L,R}$ fermions are straightforward to
obtain,
\begin{align}
\frac{\delta\mathcal{L}_{\text{Yukawa}}}{\delta(X_{\mathbf{\bar{5}}}^{L}%
)_{A}}  &  =(\bar{X}_{\mathbf{\bar{5}}}^{R})^{B}(\mathbf{\bar{M}}_{5})_{B}%
^{A}+\sqrt{2}[(\bar{X}_{\mathbf{10}}^{L,\mathrm{C}})^{AB}\mathbf{Y}_{5}%
^{L,T}+(\bar{\chi}_{\mathbf{10}}^{\mathrm{C}})^{AB}\mathbf{\bar{Y}}_{5}%
^{T}](h_{\mathbf{\bar{5}}}^{d})_{B}\;,\frac{{}}{{}}\\
\frac{\delta\mathcal{L}_{\text{Yukawa}}}{\delta(X_{\mathbf{\bar{5}}}^{R}%
)_{A}}  &  =(\bar{X}_{\mathbf{\bar{5}}}^{L})^{B}(\mathbf{\bar{M}}_{5})_{B}%
^{A}+\sqrt{2}(\bar{X}_{\mathbf{10}}^{R,\mathrm{C}})^{AB}\mathbf{Y}_{5}%
^{R,T}(h_{\mathbf{\bar{5}}}^{d})_{B}+(\bar{\psi}_{\mathbf{\bar{5}}}%
)^{B}\mathbf{\bar{N}}_{5}(H_{\mathbf{24}})_{B}^{A}\;,\frac{{}}{{}}%
\end{align}
where $(\mathbf{\bar{M}}_{5})_{B}^{A}=\mathbf{M}_{5}\delta_{B}^{A}%
+\mathbf{N}_{5}(H_{\mathbf{24}})_{B}^{A}$.

This set of coupled equations of motion can be solved iteratively. Only the
leading order in $h_{\mathbf{5}}^{u}$ and $h_{\mathbf{\bar{5}}}^{d}$ needs to
be kept since $v_{5}^{u,d}\ll v_{24}$, and it consists of only five terms%
\begin{align}
\mathcal{L}_{\text{Yukawa}}^{eff}  &  =\sqrt{2}\left(  \bar{\psi
}_{\mathbf{\bar{5}}}\right)  ^{G}(H_{\mathbf{24}})_{G}^{E}\mathbf{\bar{N}}%
_{5}(\mathbf{\bar{M}}_{5}^{-1})_{E}^{D}(h_{\mathbf{\bar{5}}}^{d\dagger}%
)^{C}\mathbf{Y}_{5}^{L\ast}(\mathbf{\bar{M}}_{10}^{-1T})_{CD}^{AB}%
\mathbf{\bar{N}}_{10}^{T}(H_{\mathbf{24}})_{B}^{F}(\chi_{\mathbf{10}%
}^{\mathrm{C}})_{AF}\frac{{}}{{}}\nonumber\\
&  \;\;\;\;+\sqrt{2}\left(  \bar{\psi}_{\mathbf{\bar{5}}}\right)
^{D}(h_{\mathbf{\bar{5}}}^{d\dagger})^{C}\mathbf{\bar{Y}}_{5}^{\ast
}(\mathbf{\bar{M}}_{10}^{-1T})_{CD}^{AB}\mathbf{\bar{N}}_{10}^{T}%
(H_{\mathbf{24}})_{B}^{E}(\chi_{\mathbf{10}}^{\mathrm{C}})_{AE}\frac{{}}{{}%
}\nonumber\\
&  \;\;\;\;-\sqrt{2}\left(  \bar{\psi}_{\mathbf{\bar{5}}}\right)
^{C}(H_{\mathbf{24}})_{C}^{B}\mathbf{\bar{N}}_{5}(\mathbf{\bar{M}}_{5}%
^{-1})_{B}^{A}(h_{\mathbf{\bar{5}}}^{d\dagger})^{D}\mathbf{\bar{Y}}_{5}^{\ast
}(\chi_{\mathbf{10}}^{\mathrm{C}})_{AD}\frac{{}}{{}}\nonumber\\
&  \;\;\;\;-\frac{1}{4}(H_{\mathbf{24}})_{K}^{D}(\bar{\chi}_{\mathbf{10}%
}^{\mathrm{C}})^{CK}\mathbf{\bar{N}}_{10}^{T}(\mathbf{\bar{M}}_{10}%
^{-1T})_{CD}^{AB}\varepsilon_{ABEFJ}(h_{\mathbf{5}}^{u})^{J}\mathbf{Y}%
_{10}^{L}(\mathbf{\bar{M}}_{10}^{-1})_{GH}^{EF}\mathbf{\bar{N}}_{10}%
(H_{\mathbf{24}})_{L}^{H}(\chi_{\mathbf{10}})^{GL}\nonumber\\
&  \;\;\;\;+\frac{1}{2}(\bar{\chi}_{\mathbf{10}}^{\mathrm{C}})^{EF}%
\varepsilon_{ABEFG}(h_{\mathbf{5}}^{u})^{G}\mathbf{\bar{Y}}_{10}%
(\mathbf{\bar{M}}_{10}^{-1})_{CD}^{AB}\mathbf{\bar{N}}_{10}(H_{\mathbf{24}%
})_{H}^{D}(\chi_{\mathbf{10}})^{CH}+h.c.\;.
\end{align}
Note that $\mathbf{Y}_{5,10}^{R}$ do not contribute at all. If $\mathbf{M}%
_{i}\gg v_{24}\mathbf{N}_{i}$, the inverse mass terms can be expanded as%
\begin{align}
\mathbf{M}_{5}\cdot\frac{1}{\mathbf{M}_{5}\delta_{B}^{A}+\mathbf{N}%
_{5}(H_{\mathbf{24}})_{B}^{A}}  &  =\delta_{B}^{A}\mathbf{1}+\mathbf{N}%
_{5}\frac{(H_{\mathbf{24}})_{B}^{A}}{\mathbf{M}_{5}}+\mathbf{N}_{5}%
\frac{(H_{\mathbf{24}})_{C}^{A}}{\mathbf{M}_{5}}\mathbf{N}_{5}%
\frac{(H_{\mathbf{24}})_{B}^{C}}{\mathbf{M}_{5}}+...\;,\\
\mathbf{M}_{10}\cdot\frac{1}{\mathbf{M}_{10}\delta_{AB}^{CD}+\mathbf{N}%
_{10}(H_{\mathbf{24}})_{AB}^{CD}}  &  =\delta_{CD}^{AB}\mathbf{1}%
+\mathbf{N}_{10}\frac{(H_{\mathbf{24}})_{CD}^{AB}}{\mathbf{M}_{10}}%
+\mathbf{N}_{10}\frac{(H_{\mathbf{24}})_{CD}^{EF}}{\mathbf{M}_{10}}%
\mathbf{N}_{10}\frac{(H_{\mathbf{24}})_{EF}^{AB}}{\mathbf{M}_{10}}+...\;.
\end{align}
To leading order in $\mathbf{M}_{i}^{-1}$, the effective interactions become:%
\begin{align}
\mathcal{L}_{\text{Yukawa}}^{eff}  &  =-\sqrt{2}(H_{\mathbf{24}})_{B}%
^{A}\left(  \bar{\psi}_{\mathbf{\bar{5}}}\right)  ^{B}\left[  \frac{1}%
{2}\mathbf{\bar{Y}}_{5}^{\ast}\frac{1}{\mathbf{M}_{10}}\mathbf{\bar{N}}%
_{10}^{T}+\mathbf{\bar{N}}_{5}\frac{1}{\mathbf{M}_{5}}\mathbf{\bar{Y}}%
_{5}^{\ast}\right]  (\chi_{\mathbf{10}}^{\mathrm{C}})_{AD}(h_{\mathbf{\bar{5}%
}}^{d\dagger})^{D}\nonumber\\
&  \;\;\;\;-\sqrt{2}\left(  \bar{\psi}_{\mathbf{\bar{5}}}\right)  ^{A}\left[
\frac{1}{2}\mathbf{\bar{Y}}_{5}^{\ast}\frac{1}{\mathbf{M}_{10}}\mathbf{\bar
{N}}_{10}^{T}\right]  (\chi_{\mathbf{10}}^{\mathrm{C}})_{AB}(H_{\mathbf{24}%
}h_{\mathbf{\bar{5}}}^{d\dagger})^{B}\nonumber\\
&  \;\;\;\;+\frac{1}{2}\varepsilon_{ABCDE}(\bar{\chi}_{\mathbf{10}%
}^{\mathrm{C}})^{AB}\left[  \mathbf{\bar{Y}}_{10}\frac{1}{\mathbf{M}_{10}%
}\mathbf{\bar{N}}_{10}\right]  (\chi_{\mathbf{10}})^{CF}(H_{\mathbf{24}}%
)_{F}^{D}(h_{\mathbf{5}}^{u})^{E}+h.c.\;.
\end{align}
The $X_{\mathbf{5}}^{L,R}$ fermions induce only the $45$-type interactions
(which includes some left-over $5$-type as $h_{\mathbf{5}}^{D}(H_{\mathbf{24}%
})_{B}^{A}$ is not traceless), while the $X_{\mathbf{10}}^{L,R}$ fermions
generate all types of effective interactions but for those of the form
$\varepsilon_{ABCDE}(\bar{\chi}_{\mathbf{10}}^{\mathrm{C}})^{AB}%
(\chi_{\mathbf{10}})^{CD}(H_{\mathbf{24}}h_{\mathbf{5}})^{E}$, because the
$H_{\mathbf{24}}$ indices are all used to contract those of the $\chi
_{\mathbf{10}}$ and never ends up coupled to that of $h_{\mathbf{5}}^{u}$ or
$h_{\mathbf{\bar{5}}}^{d}$.

The general expression does not permit to easily extract the contributions to
the fermion Yukawa couplings after the SSB. To this end, we set $[$no
summation on $A,B]$%
\begin{align}
\mathbf{M}_{5}\delta_{B}^{A}+\mathbf{N}_{5}(H_{\mathbf{24}})_{B}^{A}  &
\rightarrow(\mathbf{M}_{5}+\mathbf{N}_{5}v_{24}^{A})\times\delta_{B}%
^{A}\frac{{}}{{}}\;,\\
\mathbf{M}_{10}\delta_{AB}^{CD}+\mathbf{N}_{10}(H_{\mathbf{24}})_{AB}^{CD}  &
\rightarrow(\mathbf{M}_{10}+\mathbf{N}_{10}(v_{24}^{A}+v_{24}^{B}%
)/2)\times\delta_{AB}^{CD}\;.
\end{align}
Plugging this in the general expression, setting $v_{24}=(1,1,1,-3/2,-3/2)$,
and denoting $\mathbf{F}_{\mathbf{R}}^{\alpha}=\alpha v_{24}\mathbf{N}%
_{\mathbf{R}}^{M}\cdot(\mathbf{M}_{\mathbf{R}}+\alpha v_{24}\mathbf{N}%
_{\mathbf{R}})^{-1}$, we find the effective Yukawa interactions of
Eq.~(\ref{EffectiveSU5Yuk}).

\end{document}